\documentclass[11pt]{article}
\usepackage{amsmath}
\usepackage{amssymb}
\usepackage{slashed}
\usepackage{graphicx}
\usepackage{graphics}
\usepackage{epsfig}
\usepackage{soul}
\usepackage{hyperref}
\usepackage{upgreek}
\usepackage{authblk}
\usepackage{blindtext}
\usepackage[section]{placeins}
\usepackage{mathrsfs}
\usepackage{dsfont}

\usepackage{graphicx}

\usepackage[normalem]{ulem}

\usepackage{dcolumn}
\usepackage{bm}

\usepackage{color}
\usepackage[usenames,dvipsnames,svgnames,table]{xcolor}
\usepackage{slashed}
\usepackage{empheq}
\newcommand{\bea}{\begin{eqnarray}}
\newcommand{\eea}{\end{eqnarray}}
\newcommand{\be}{\begin{equation}}
\newcommand{\ee}{\end{equation}}

\usepackage[latin1]{inputenc} 
 \author[1]{Astrid Eichhorn \thanks{a.eichhorn@thphys.uni-heidelberg.de}}

\affil[1]{Institut f\"ur Theoretische
  Physik, Universit\"at Heidelberg, Philosophenweg 16, 69120
  Heidelberg, Germany}
\title{An asymptotically safe guide to quantum gravity and matter}

\begin{document}

\maketitle

\begin{abstract}
Asymptotic safety generalizes asymptotic freedom and could contribute to understanding physics beyond the Standard Model. It is a candidate scenario to provide an ultraviolet extension for the effective quantum field theory of gravity through an interacting fixed point of the Renormalization Group. Recently, asymptotic safety has been established  in specific gauge-Yukawa models in four dimensions in perturbation theory, providing a starting point for asymptotically safe model building. Moreover, an asymptotically safe fixed point might even be induced in the Standard Model under the impact of quantum fluctuations of gravity in the vicinity of the Planck scale. This review contains an overview of the key concepts of asymptotic safety, its application to matter and gravity models,  exploring potential phenomenological implications and highlighting open questions.
\end{abstract}

\section{Invitation to asymptotic safety}

\begin{figure}[!t]
 \centering
 \includegraphics[width=0.5\linewidth]{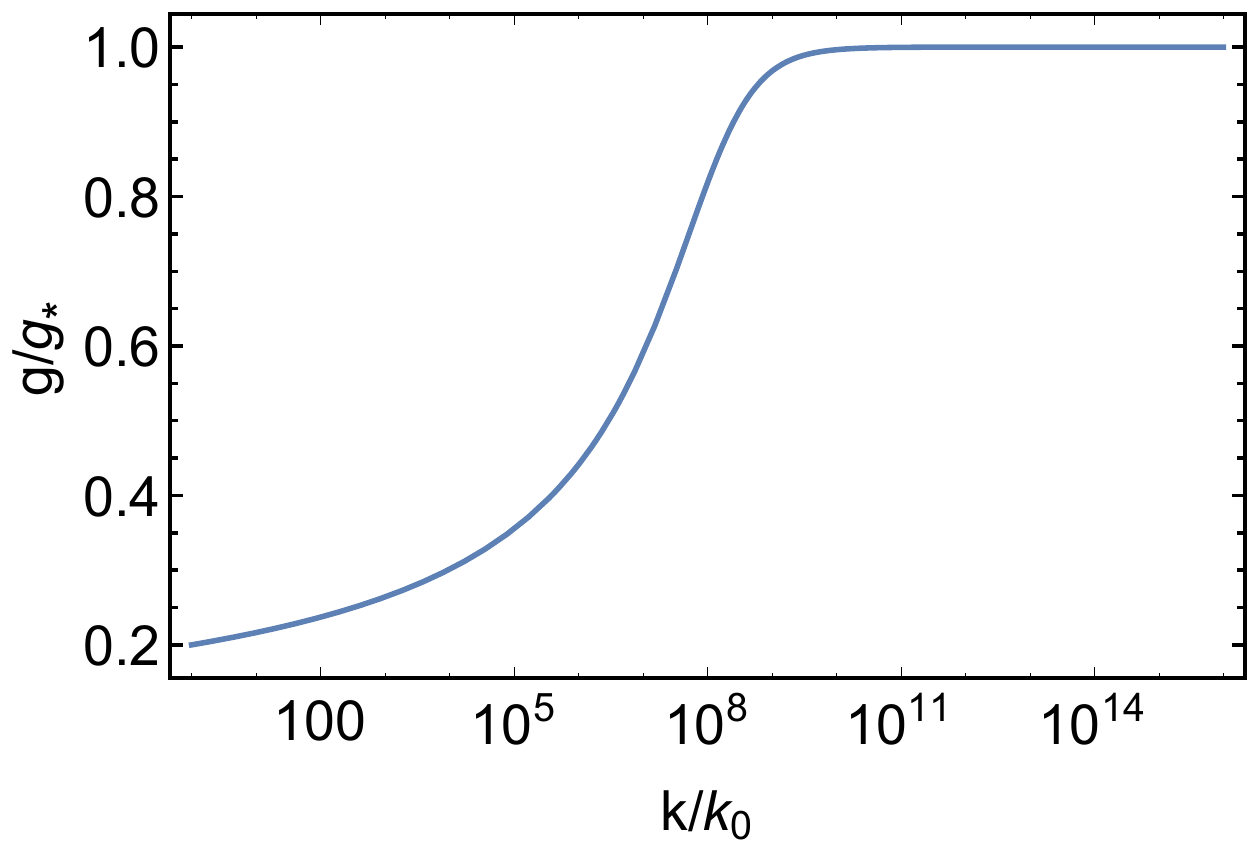}
 \caption{\label{fig:ASschem}Schematic RG flow for an asymptotically safe coupling. Beyond the transition scale at $k/k_0 \approx 10^9$, (approximate) scale-invariance is realized; full scale-invariance is realized asymptotically at $k/k_0 \rightarrow \infty$.}
 \end{figure}

Asymptotic safety \cite{Weinberg:1980gg} is a quantum-field theoretic paradigm providing an ultraviolet (UV) extension or completion for effective field theories. The high-momentum regime of an asymptotically safe theory is scale invariant, cf.~Fig.~\ref{fig:ASschem}. It is governed by a fixed point of the Renormalization Group (RG) flow of couplings.
As such, asymptotic safety is an example of a fruitful transfer of ideas from statistical physics to high-energy physics: In the former, interacting RG fixed points provide universality classes for continuous phase transitions \cite{Wilson:1971dc,ZinnJustin:2002ru}, in the latter these generalize asymptotic freedom to a scale-invariant UV completion with residual interactions. This paradigm is being explored for physics beyond the Standard Model in several promising ways.
Following the discovery of perturbative asymptotic safety in weakly-coupled gauge-Yukawa models in four dimensions \cite{Litim:2014uca}, the search for asymptotically safe extensions of the Standard Model with new degrees of freedom close to the electroweak scale is ongoing.
Mechanisms for asymptotic safety also exist in nonrenormalizable settings, making it a candidate paradigm for quantum gravity \cite{Weinberg:1980gg,Reuter:1996cp}. 
After the discovery of the Higgs boson \cite{Aad:2012tfa,Chatrchyan:2012xdj}, we know that the Standard Model can consistently be extended up to the Planck scale \cite{Bezrukov:2012sa,Buttazzo:2013uya,Bezrukov:2014ina}. Hence,  the interplay of the Standard Model with quantum fluctuations of gravity within a quantum field theoretic setting is under active exploration.

This review aims at providing an introduction to asymptotic safety for non-experts, highlighting mechanisms that generate asymptotically safe physics, explaining how these could play a role in settings relevant for high-energy physics and discussing open questions of (potentially) asymptotically safe models. An extensive bibliography is intended to serve as a guide to further reading, providing more comprehensive and in-depth answers to many points only touched upon briefly here.

\section{Asymptotic safety - the key idea}\label{sec:ASkey}
Quantum fluctuations induce a momentum-scale dependence in the couplings of a model, breaking scale invariance even in classically scale-invariant models. Scale invariance is restored at RG fixed points. These can be non-interacting, in which case the theory is asymptotically free, or interacting in at least one of the couplings, in which case the theory is asymptotically safe.
Both fixed points underlie theories that are fundamental in a Wilsonian sense: For a theory that is discretized, e.g., on a lattice, an RG fixed point guarantees that a continuum limit exists. Scale-invariance protects the running couplings in a model from Landau poles which can signal a breakdown of a description of an interacting system by this model because of triviality 
\footnote{Models affected by the triviality problem can only hold up to arbitrarily high momentum scales if the coupling vanishes at all scales,  rendering the models noninteracting, or trivial. Establishing triviality requires going beyond perturbation theory. Nonetheless, an intuitive understanding of the problem can be gained from perturbation theory, e.g., in scalar $\lambda\, \phi^4$ theory in four dimensions. The one-loop beta function for the quartic self-interaction $\lambda$ is $\beta_{\lambda} = \# \lambda^2$, where $\#>0$ holds. Integrating the beta function leads to a logarithmic divergence. Pushing the scale $\Lambda$ of the divergence (the Landau pole) to infinity requires $\lambda_0=\lambda(k_0)=0$, since $\Lambda/k_0 =e^{\frac{1}{\# \lambda_0}}$.}.
 Hence, the introduction of new physics is  one viable theoretical option instead of a necessity.
 
Scale-invariance requires a fixed point in the dimensionless couplings $g_i$, obtained from their dimensionful counterparts $\bar{g}_i$ with canonical dimension $d_{\bar{g}_i}$ by multiplication with an appropriate power of the RG scale $k$
\be
g_i(k)= \bar{g}_i(k)\, k^{-d_{\bar{g}_i}}\,.
\ee
A scale-invariant point is a zero of all beta functions, 
\be
\beta_{g_i}= k\, \partial_k\, g_i(k) = 0\, \quad {\rm at} \quad g_{i}= g_{i\, \ast}.
\ee
 Then, dimensionful couplings 
\footnote{\cite{Weinberg:1980gg} motivates the focus on dimensionless couplings $g_i$ instead of their dimensionful counterparts $\bar{g}_i$ by requiring finiteness of observables. Measurable quantities at some energy scale $E$, e.g., a scattering cross-section $\sigma$, can be written as  $\sigma =E^{\#}f(g_i)$, where $\#$ is the canonical dimension of $\sigma$, multiplied by a function of the \emph{dimensionless} couplings $g_i$ that enter. Herein the RG scale is equated to a physical energy scale. If the dimensionless couplings diverge at a finite energy scale, this typically entails divergences in physical quantities.}
 scale with their canonical dimensionality, i.e., $\bar{g}_i(k)\sim k^{d_{\bar{g}_i}}$, since $g_i(k) = g_{i\, \ast}= \rm const$ in a scale-invariant regime. This must hold for all couplings in the infinite-dimensional theory space, spanned by all interactions allowed by symmetries, including higher-order, i.e., canonically irrelevant interactions. Quantum fluctuations generically generate all interactions, as familiar from effective field theories. Moreover, there is no a priori physical argument to exclude higher-order terms from the dynamics. The restriction to perturbatively renormalizable terms that is commonly assumed is actually an automatic consequence of the universality class of the Gaussian, i.e., free fixed point which renders higher-order terms irrelevant for perturbative low-energy physics.

\subsection{Predictivity in the infinite-dimensional space of couplings}

\begin{figure}[!t]
\includegraphics[width=0.48\linewidth,clip=true, trim=11cm 4cm 12.8cm 12cm]{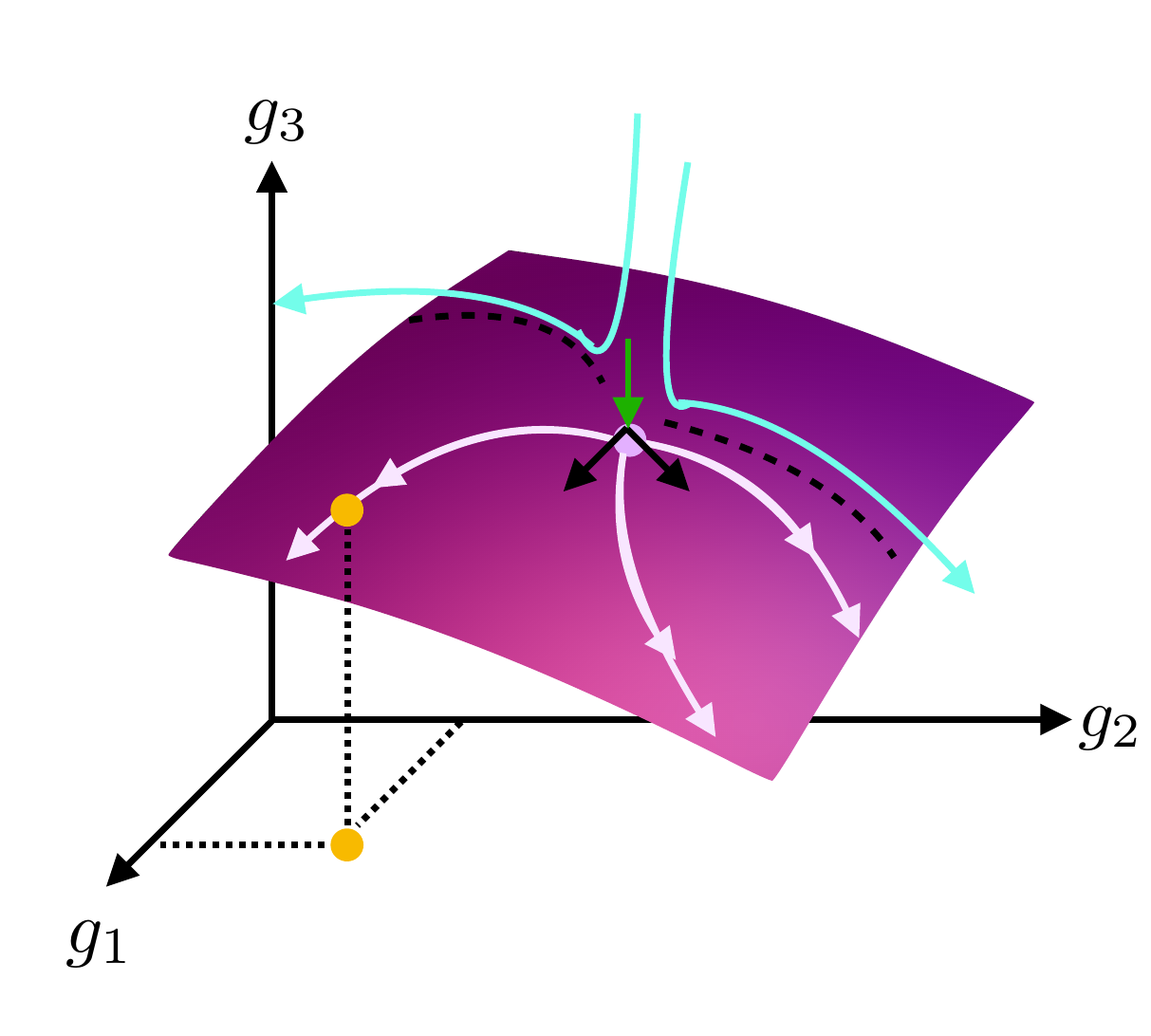}
\includegraphics[width=0.48\linewidth,clip=true,trim=1cm 4cm 12cm 0cm]{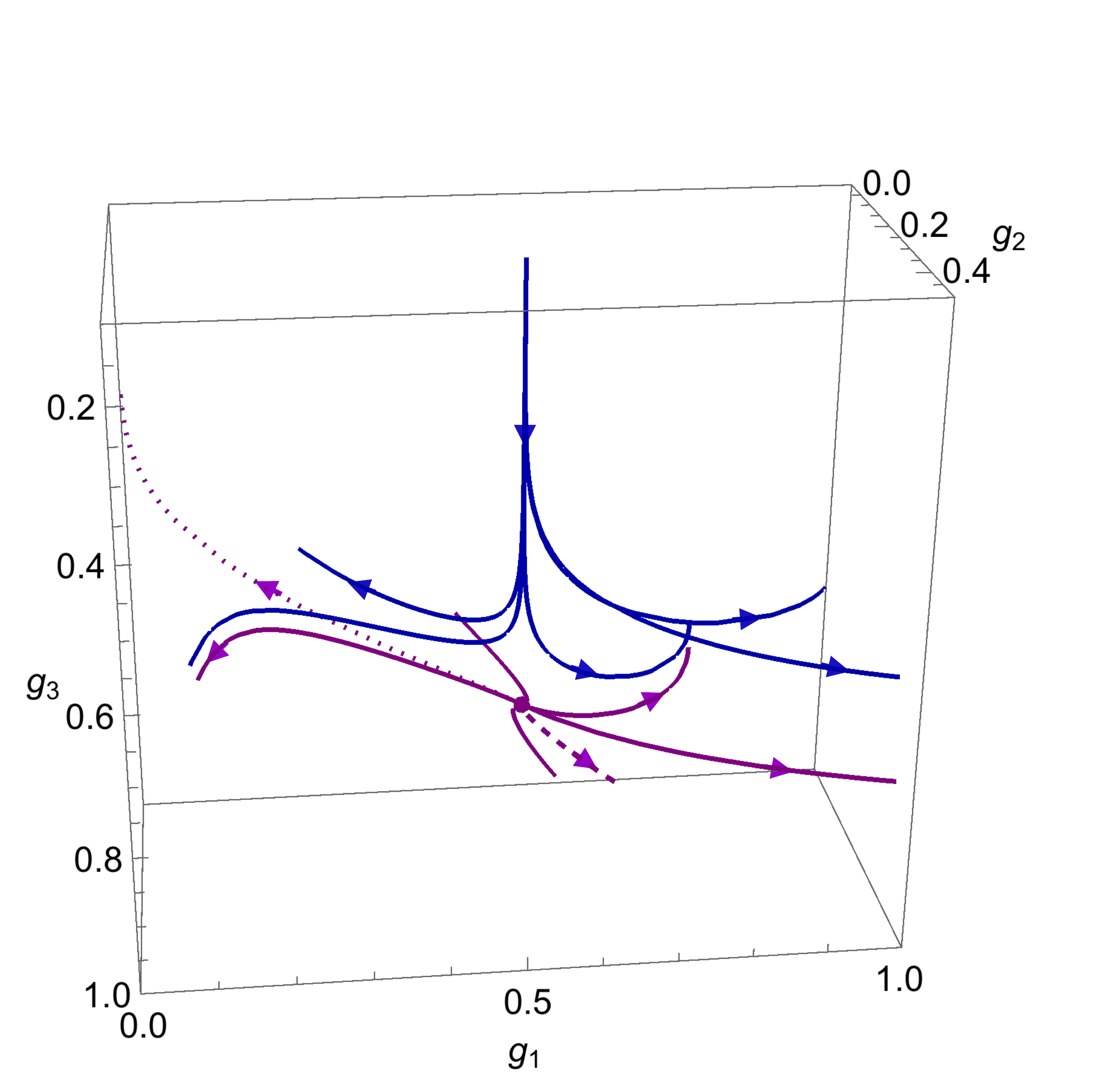}
\caption{\label{fig:FPattractive} 
Left panel: Illustration of a fixed point (light purple dot) with its UV critical hypersurface (purple). RG trajectories starting off the critical hypersurface (teal) are pulled towards the fixed point along the irrelevant direction (roughly aligned with $g_3$), before the IR repulsive directions $g_1$ and $g_2$ kick in and drive the flow away from the fixed point. The linearized flow is indicated by the black (relevant directions) and green (irrelevant direction) arrows.
Right panel: A fixed point with two relevant directions in the space of three couplings $g_{1,2,3}$, where $g_3$ corresponds to the IR attractive direction. The purple trajectories emanate from the fixed point, and $g_{2,3}$ fully determine the deviation from scale invariance. The arrows indicate the RG flow towards the IR. The corresponding beta functions of one canonically marginal, relevant and irrelevant interaction are given by 
$\beta_{g_1}= 2 g_1 -3g_1^2-3 g_1^2\, g_2$, $\beta_{g_2}=-2g_2\, +2g_1- 3g_1\, g_2$ and $\beta_{g_3}= -g_1\, g_3+g_3^3$.
}
\end{figure}

The main consequence of an RG fixed point is not that it provides a fundamental theory -- after all, experiments are  limited to finite scales -- but that it generates universal predictions for low-energy physics. It imposes relations between the couplings encoding the location of the UV-critical surface of the fixed point. This hypersurface is spanned by all couplings along which RG trajectories emanate from the fixed point as one lowers $k$ towards the infrared (IR). The corresponding \emph{relevant} directions  parameterize the deviation from scale invariance. They constitute  free parameters, as a range of values of relevant couplings can be reached along different trajectories emanating from the fixed point, cf.~Fig.~\ref{fig:FPattractive}. It can be more intuitive to understand that a free parameter is associated to such a direction, as IR-repulsiveness equals UV-attractivity. Irrespective of its IR value, a UV-attractive coupling reaches the fixed point at high scales as one reverses the flow towards the UV. (Nevertheless, recall that although we measure physics at low energies and try to extrapolate towards viable UV physics, nature works the other way: IR physics emerges as a consequence of UV physics.) \\
Towards the IR, the irrelevant, i.e., IR-attractive directions are automatically pulled towards the fixed point, cf.~Fig.~\ref{fig:FPattractive}. Accordingly, no free parameter is associated to them -- this is the universality-generating property of an RG fixed point: Initializing the RG flow at some scale $k_0$, the flow maps a UV interval of values for an irrelevant coupling at $k_0$ to a much smaller IR interval. The latter shrinks to zero as one takes $k_0 \rightarrow \infty$. As a one-coupling example with an IR attractive fixed point, consider
\be
\beta_{g}= g(g- g_{\ast}),
\ee
with the solution
\be
g(k) = \frac{g_{\ast}}{1+\left(\frac{k}{k_0}\right)^{g_{\ast}} \left(\frac{g_{\ast}}{g_0}-1 \right)},
\ee
where $g(k_0)=g_0$. As the initial scale $k_0 \rightarrow \infty$, $g(k) \rightarrow g_{\ast}$. For a finite $k_0$, $g(k)-g_{\ast}$ goes to zero as $k/k_0 \rightarrow 0$.
For a trajectory that emanates from the fixed point, there is no freedom of choice choice for the value for an irrelevant direction: the fixed-point requirement restricts the flow to lie within the critical hypersurface, resulting in completely determined values for the irrelevant directions.
For instance, at the free fixed point,  higher-order couplings do not play a role in the IR: the RG flow drives them towards zero for all perturbative initial conditions in the UV.  This generates universality and independence of the IR physics from the UV physics in all but the (marginally) relevant couplings.

\begin{figure}[!t]
\includegraphics[width=0.45\linewidth]{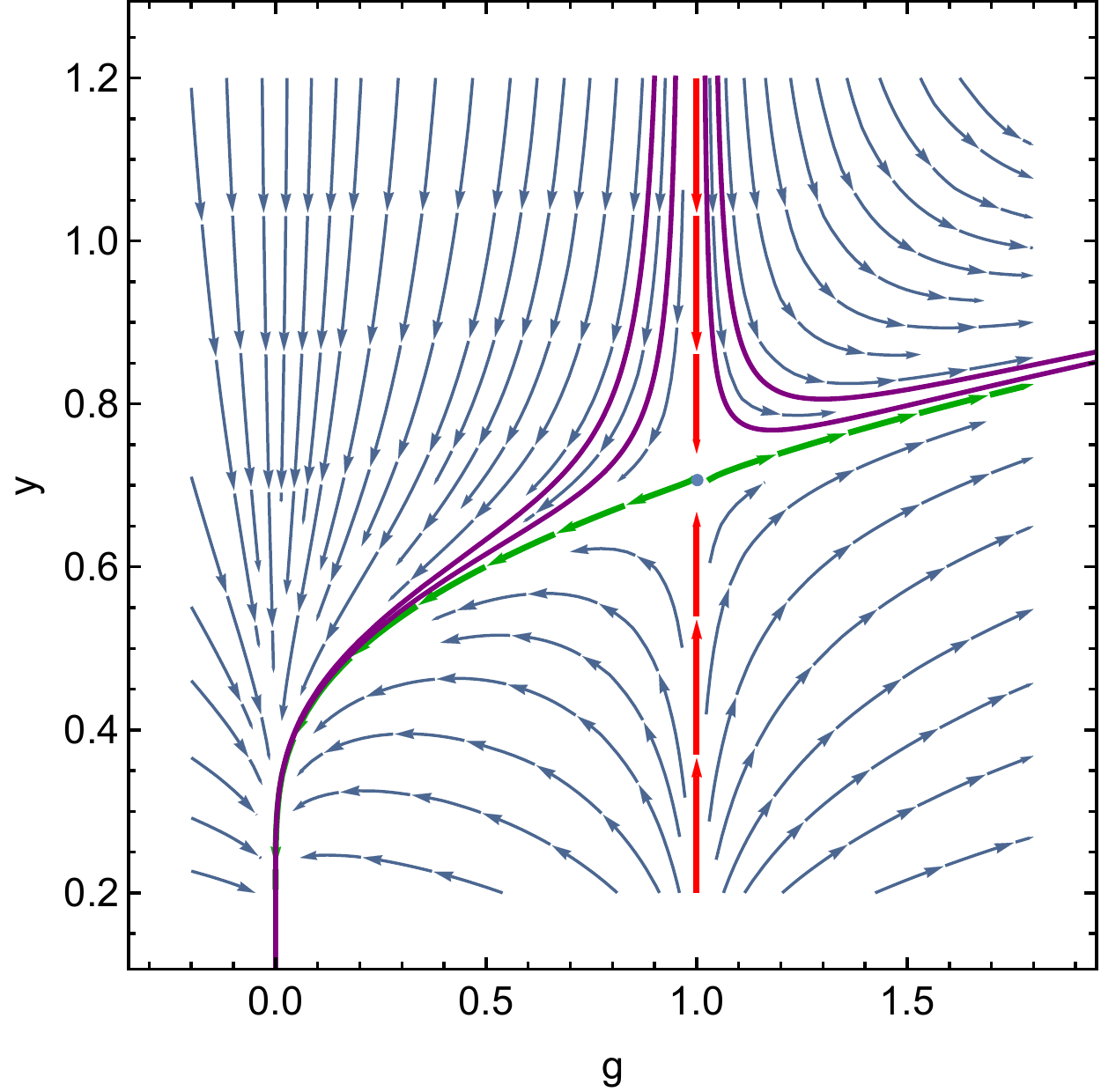}\quad \includegraphics[width=0.45\linewidth]{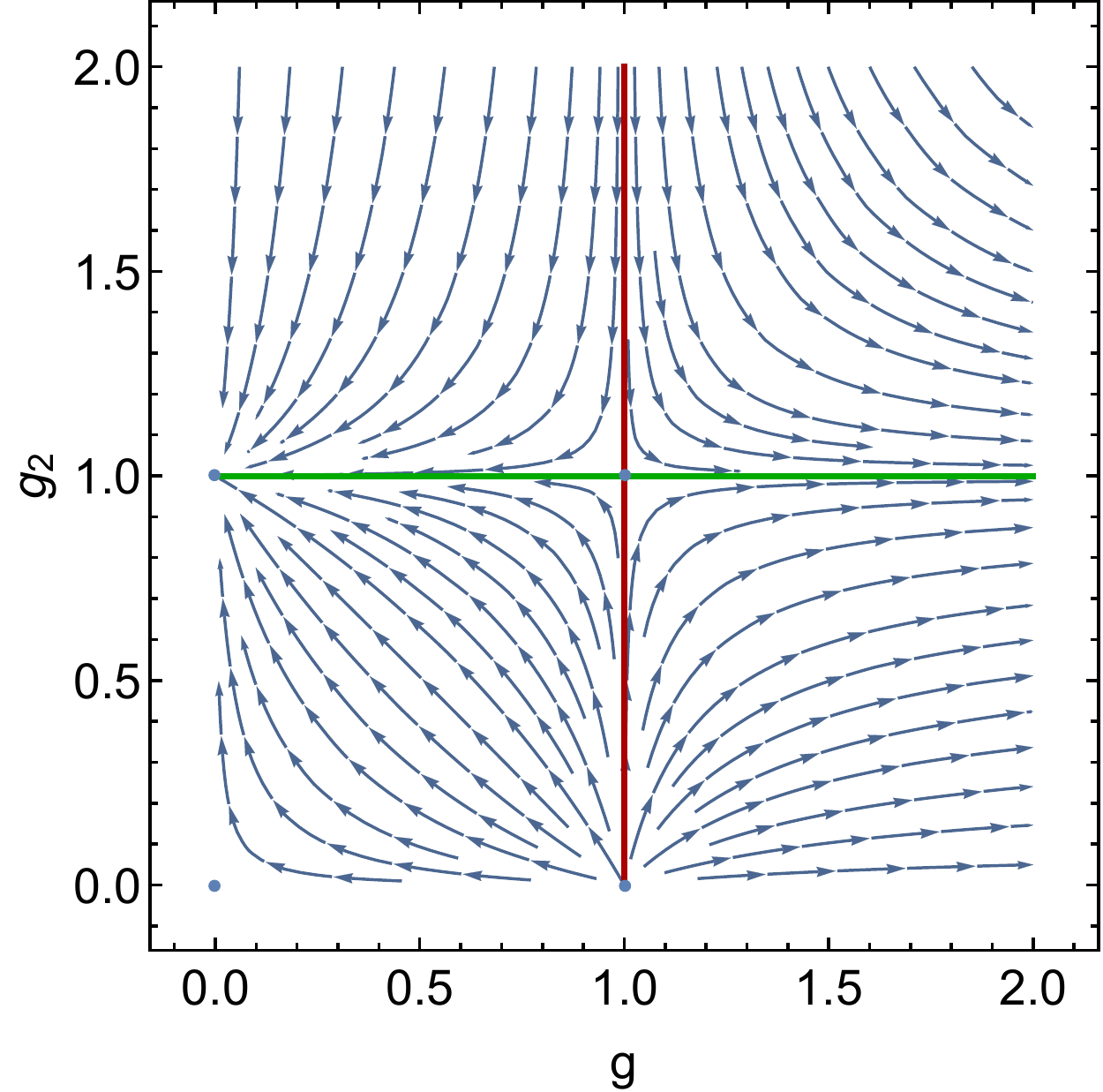}
\caption{\label{fig:IR_UV_critsurf} Left panel: The beta functions $\beta_{g}=2g-2g^2$ and $\beta_{y}= -g\, y +2 y^3$ feature a fixed point at $g=1,\, y=1/\sqrt{2}$ that has one UV attractive and one IR attractive direction. The UV critical surface is indicated in green, the IR critical surface in red. The RG flow towards the IR is attracted towards the UV critical surface, such that the relation between $g$ and $y$ that parameterizes the UV critical surface is approximately realized also for trajectories (in purple) that start off the UV critical surface. Right panel: The flow described by $\beta_{g}=2g-2g^2$ and $\beta_{g_2}=-2g_2+2g_2^2$ features a fixed point at $g=g_2=1$, which is IR attractive in $g_2$ and where the UV critical surface has no curvature. Therefore $g_2(k)=1$ for the trajectories emanating from this fixed point.}
\end{figure}

To determine the set of IR-repulsive (= UV attractive) directions, it suffices to examine the linearized flow about the fixed point 
\footnote{To determine the basin of attraction of the fixed point, one numerically integrates the RG flow to generate full trajectories.} 
at $\vec{g} = \vec{g}_{\ast}$,
\be
\beta_{g_i} = \sum_j\frac{\partial\beta_{g_i}}{\partial g_j}\Big|_{\vec{g}= \vec{g}_{\ast}} \left(g_j-g_{j\, \ast} \right)+ \mathcal{O} \left(g_j-g_{j\, \ast} \right)^2.\label{eq:linflow}
\ee
In terms of the critical exponents\footnote{The opposite sign convention is sometimes used in the literature.}
\be
\theta_I =-{\rm eig} \mathcal{M}_{ij}= -{\rm eig} \frac{\partial\beta_{g_i}}{\partial g_j}\Big|_{\vec{g}= \vec{g}_{\ast}},
\ee
and corresponding (right) eigenvectors $V^I$, the solution to Eq.~\eqref{eq:linflow} is 
\be
g_i(k) = g_{i\, \ast} + \sum_I c_I\, V_i^I\, \left(\frac{k}{k_0} \right)^{-\theta_I}.\label{eq:linflowsol}
\ee
$k_0$ is an arbitrary reference scale and $c_I$ are constants of integration. Typically, the set of couplings $\vec{g}$ does not diagonalize the stability matrix $\mathcal{M}_{ij}$ at $\vec{g} = \vec{g}_{\ast}$ and the eigenvectors $V^I$ are  superpositions. As the stability matrix need not be symmetric, the eigenvalues need not be real. Their imaginary part results in a spiralling behavior of the flow in the vicinity of the fixed point, where the real part determines whether the spiralling is inwards or outwards. To determine the set of free parameters, it therefore suffices to consider the real parts. For the following discussion we will thus assume that the eigenvalues are real.
For $\theta_I>0$, the corresponding eigenvector $V^I$ constitutes an IR repulsive direction: Towards the IR, the distance to the fixed-point regime grows, and the IR values of couplings appearing in $V^I$ depend on $c_I$. Fixing this free parameter  requires experimental input. Accordingly, predictivity requires a finite number of directions with $\theta_I>0$.\\
In contrast, for $\theta_I<0$, the IR values of couplings are independent of the corresponding $c_I$, cf.~Eq.~\eqref{eq:linflowsol}: For $\theta_I<0$, any deviation from the fixed-point value in $V^I$ is washed out by the RG flow to the IR.  Once a choice of coordinates in theory space is made, the critical hypersurface typically exhibits curvature.
If the critical hypersurface had no curvature, the values of irrelevant couplings would be constant. Curvature of the critical hypersurface generates a  scale dependence which is completely fixed once the values of all relevant couplings are specified, cf.~Fig.~\ref{fig:FPattractive}. 
The values of the corresponding irrelevant couplings depend on the scale, but not independently of the relevant couplings, cf.~green trajectory in~Fig.~\ref{fig:IR_UV_critsurf}.

For asymptotic safety, the finite contribution to the $\theta_I$  due to residual interactions at $\vec{g}_{\ast}$ shifts the critical exponents away from the canonical dimensions of couplings, e.g.,
\be
\theta_i= - \frac{\partial \beta_i}{\partial g_i}\Big|_{g_i=g_{i\,\ast}} = - \frac{\partial }{\partial g_i}(\partial_t\, \bar{g}_i\, k^{-d_{\bar{g}_i}})\Big|_{g_i=g_{i\,\ast}} = d_{\bar{g}_i}\, g_{i\,\ast}+ \eta(g_{i\,\ast}^2),
\ee
for a coupling $g_i$ that is an eigendirection of $\mathcal{M}_{ij}$.
This can enhance the predictive power of asymptotically safe over asymptotically free models.

The interpretation of asymptotic safety as a way of imposing predictivity on a model specified by its field content and symmetries is crucial in the context of quantum gravity.
The simplest interpretation of the Planck scale suggests that it acts as a minimal length, inducing discreteness for quantum-gravity models at the kinematical level. This might suggest that one need not search for a continuum limit in quantum gravity.  Yet, by requiring a continuum limit one restricts the dynamics to a trajectory within the critical surface, leaving just a finite number of free parameters to determine the dynamics \emph{at all scales}. In the presence of an explicit cutoff scale,  the microscopic dynamics might be defined anywhere in the theory space, requiring specification of an infinite number of couplings for the UV dynamics, see also \cite{Eichhorn:2017bwe}. Similarly, predictivity \emph{at high scales} breaks down in effective field theories. Moreover, \emph{physical} discreteness can arise in quantum gravity even in a continuum theory, through the \emph{dynamical} emergence of a scale, see, e.g., \cite{Reuter:2005bb,Percacci:2010af}, or through discreteness in the spectra of operators \cite{Rovelli:1994ge,Ashtekar:1996eg}. 

\subsection{Asymptotic safety in a nutshell}
The development of the Standard Model was based on the principle of renormalizability. This is one way of implementing predictivity, i.e., constructing a low-energy theory with a finite number of free parameters.Yet, as, e.g., $\phi^4$ theory in 4 dimensions highlights, a perturbatively renormalizable theory is not guaranteed to exist as a fundamental theory in the Wilsonian sense, due to the triviality problem. Analogously, the Standard Model is actually expected to be an effective low-energy theory. Asymptotic safety is a paradigm that combines the requirement of predictivity with the possibility of obtaining a fundamental theory through an RG fixed point at high momenta with a finite number of relevant directions. The fixed point ensures nonperturbative renormalizability, while the finite dimensionality of the critical hypersurface guarantees predictivity of the model.

\subsection{Non-fundamental asymptotic safety}\label{sec:ASnonfund}

Instead of providing a ``fundamental'' UV completion, asymptotic safety might serve as one step forward in our understanding of microscopic physics, with more fundamental physics to be discovered beyond. While providing a UV completion for some RG trajectories, a fixed point can simultaneously act as an IR attractor for a more fundamental description. This follows, as a fixed point's UV repulsive directions correspond to its IR attractive directions. 
Hence, it is a misconception that a fixed point is either UV or IR - whether trajectories emanate from it in the UV, or approach it in the IR depends on the \emph{initial conditions} for the RG flow. Given \emph{two} fixed points connected by an RG trajectory, the distinction into a UV and an IR fixed point (which is also expected to satisfy the a-theorem \cite{Cardy:1988cwa}) follows from the trajectory.

For specificity, assume that a cutoff scale $k_{\rm UV}$ exists, such that for $k>k_{\rm UV}$ a (quasilocal) quantum field theoretic description is impossible or requires additional fields and/or symmetries. At $k\leq k_{\rm UV}$, the dynamics can be described in the asymptotically safe theory space. Initial conditions for the RG flow are determined by the underlying fundamental model at $k=k_{\rm UV}$. It they lie close to or on the \emph{IR-critical surface} of a fixed point, the flow is attracted towards the fixed point along its IR-attractive directions. The flow is actually driven towards the \emph{UV-critical surface}, cf.~purple trajectories in Fig.~\ref{fig:IR_UV_critsurf}. Trajectories can even spend a large amount of RG ``time" close to the fixed point. At $k_{\rm trans}<k_{\rm UV}$ the effect of the IR-repulsive directions kicks in and the flow is driven away from the fixed point along its IR-repulsive directions.  This trajectory will result in IR-values of couplings close to those of a ``true" fixed-point trajectory, cf.~Fig.~\ref{fig:IR_UV_critsurf}, see \cite{Percacci:2010af}. The above is nothing but a detailed account of how a fixed point generates IR universality. Thus, asymptotically safe fixed points could generate universal IR predictions, even in the presence of $k_{\rm UV}$.

\subsection{Mechanisms for and selected examples of asymptotic safety}
A special case of an RG fixed point is that of an asymptotically free one. To generate it, antiscreening contributions have to dominate in the beta function of the respective coupling. In contrast, asymptotic safety is generated by several different mechanisms and can be realized both in the perturbative and the nonperturbative regime, i.e., with near-Gaussian or  far-from-Gaussian critical exponents. As a second key difference, an interacting fixed point allows to combine finite, predictable IR values of couplings with UV completeness. For the free fixed point, finite IR values typically require the corresponding coupling to be an IR repulsive direction, i.e., relevant. This negates the possibility to predict the value of the coupling which remains a free parameter based on the free fixed point alone. (Of course, an interacting fixed point can dominate the flow in the IR, at which the coupling in question could be IR attractive. In this case it is again the universality class of the interacting fixed point which provides a prediction for a finite value of a coupling.)

\subsubsection{Canonical scaling versus quantum effects}\label{sec:canvsqm}

This mechanism is available for couplings which are asymptotically free in their critical dimension $d_{\rm crit}$, where they are dimensionless, i.e., their one-loop beta function is given by
\be
\beta_{g_i}\Big|_{d=d_{\rm crit}}= \beta_1\, g_i^{\#},
\ee
with $\beta_1<0$ and $\#=2,3$.
In $d= d_{\rm crit}+\epsilon$, the coupling is dimensionful, $g_i = \bar{g}_i\, k^{c\epsilon}$, where $c>0$ depends on the coupling under consideration. For $\epsilon \ll 1$, the one-loop beta function reads
 \be
 \beta_{g_i}\Big|_{d=d_{\rm crit}+\epsilon} = c\, \epsilon\, g_i+ \beta_1\, g_i^{\#}.
 \ee
An interacting fixed point lies at
 \be
 g_i^{\ast} = \left(-\frac{c\,\epsilon}{\beta_1}\right)^{1/(\#-1)}.
 \ee
 This mechanism is realized in Yang-Mills theory in $d=4+\epsilon$ \cite{Peskin:1980ay}, nonlinear sigma models in $d=2+\epsilon$ \cite{Polyakov:1975rr,Bardeen:1976zh,Friedan:1980jf,Higashijima:2003rp,Codello:2008qq,Fabbrichesi:2010xy}
 and the Gross-Neveu model in $d=2+\epsilon$ \cite{Gawedzki:1985ed,Kikukawa:1989fw,deCalan:1991km,He:1991by,Hands:1992be,Braun:2010tt}. 
 
\begin{figure}[!t]
 \includegraphics[width=0.45\linewidth]{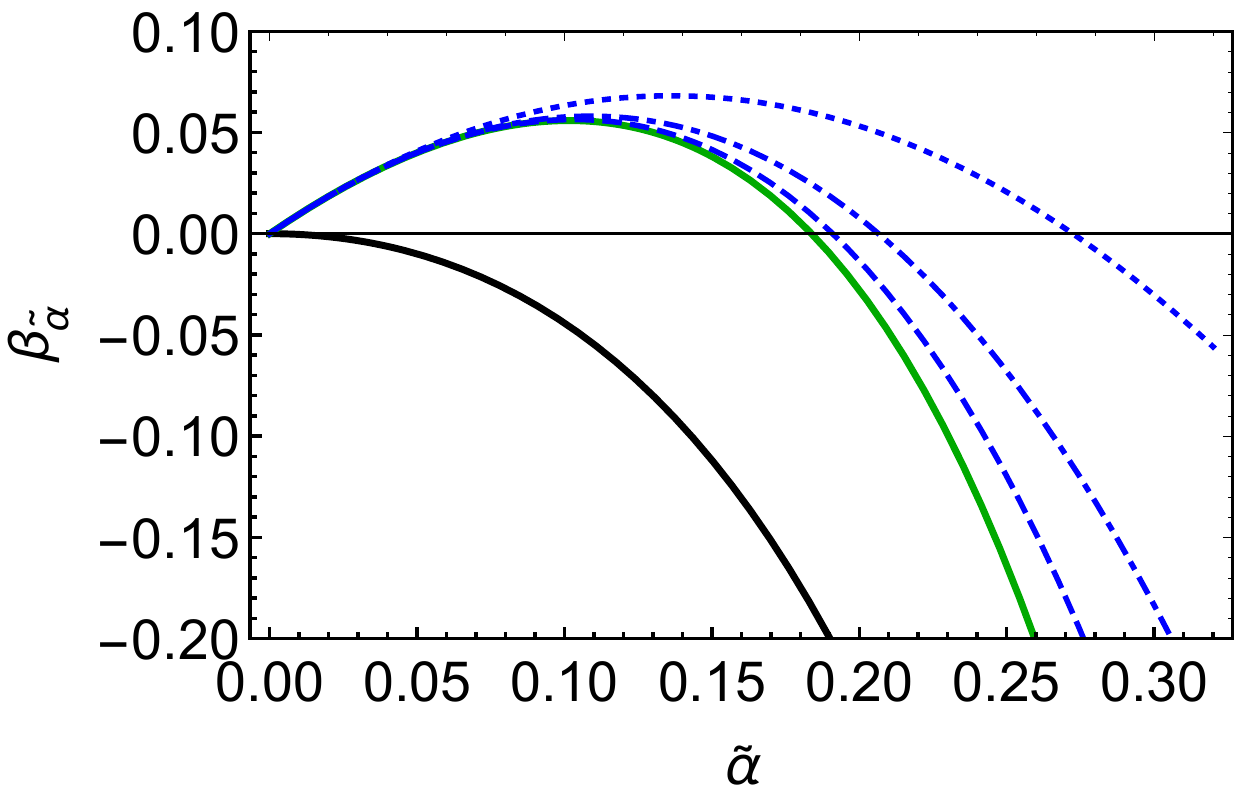}\quad\includegraphics[width=0.45\linewidth]{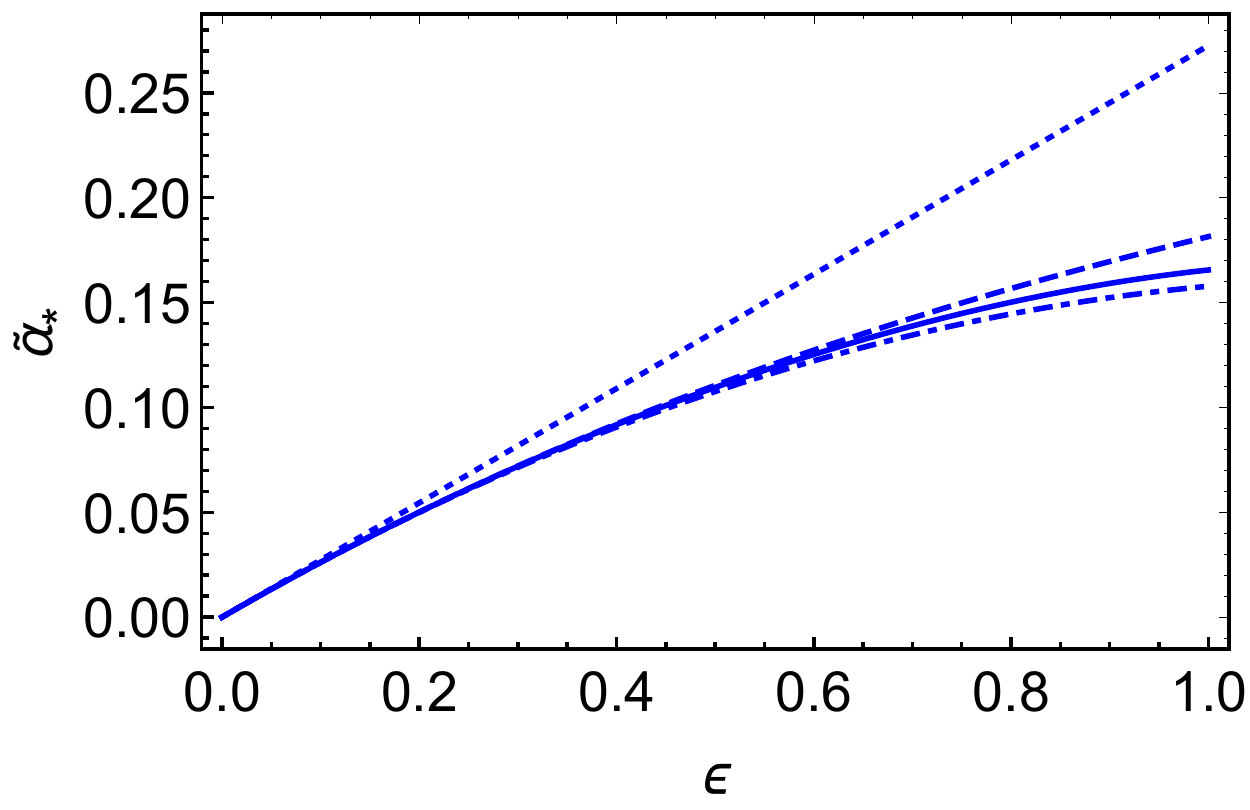}\\
 \includegraphics[width=0.45\linewidth]{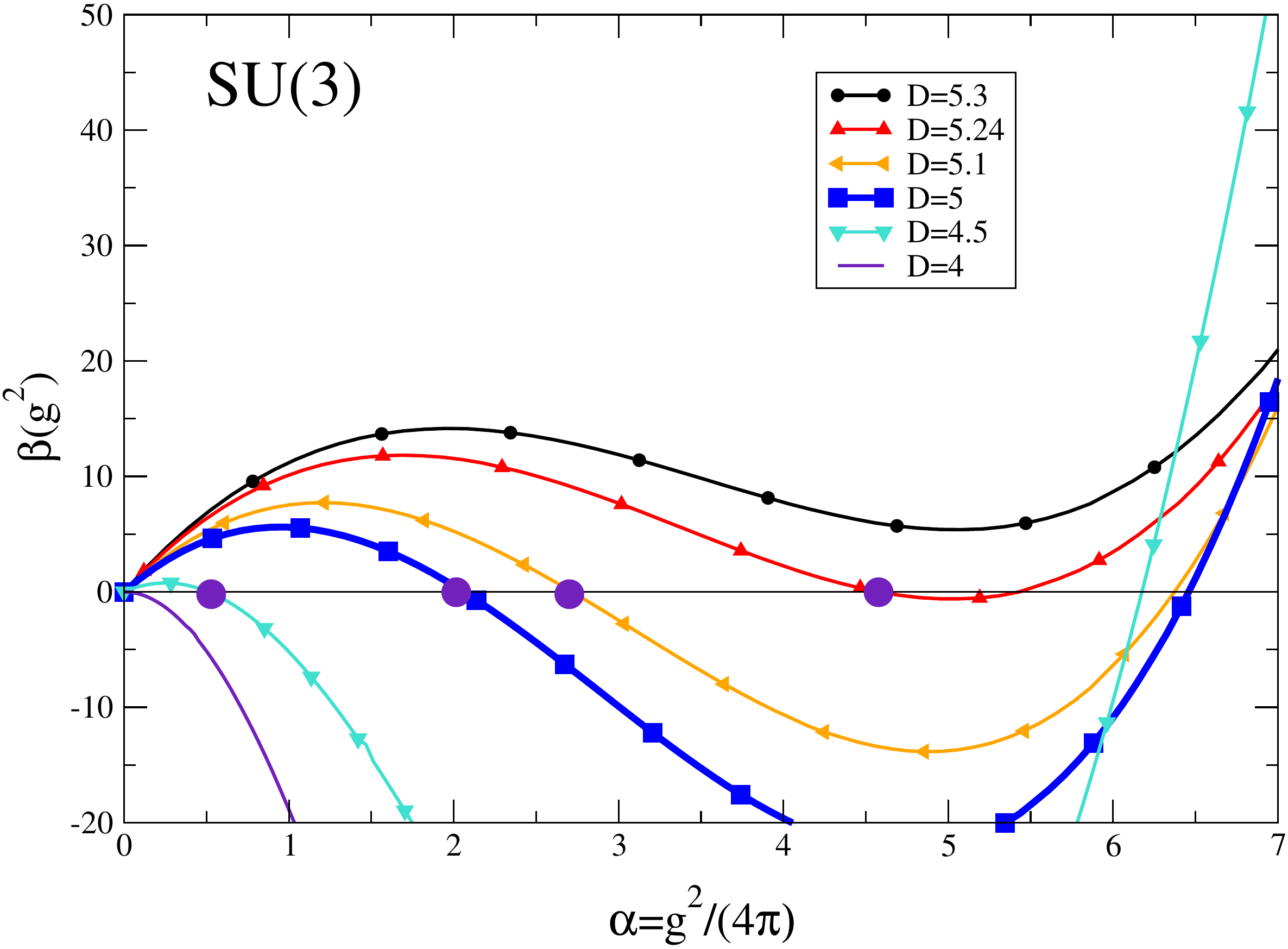}
 \caption{\label{fig:YMeps} Left upper panel: Based on results in \cite{Morris:2004mg}, the beta function for SU(3) Yang-Mills theory in the epsilon expansion for $\tilde{\alpha}= \frac{6}{(4\pi)^2}g^2$ for $\epsilon=0$ (black line), $\epsilon=1$ at one loop (dotted), two loops (dot-dashed), three loops (dashed) and four loops (green). Right upper panel: Fixed-point value for $\alpha$ as a function of $\epsilon$ up to $\epsilon$ (dotted), $\epsilon^2$ (dot-dashed), $\epsilon^3$ (dashed) and $\epsilon^4$ (continuous) emerge from the free fixed point at $d\rightarrow d_{\rm crit}$, i.e., $\epsilon \rightarrow 0$. Lower panel: Results from the FRG calculation taken from \cite{Gies:2003ic} for $\alpha=\frac{g^2}{4\pi}$.}
 \end{figure}

 For Yang-Mills theory, the $\epsilon$-expansion has been extended up to fourth order, indicating a fixed point in $d=5$ \cite{Morris:2004mg}, cf.~Fig.~\ref{fig:YMeps}, corroborating functional RG results \cite{Gies:2003ic}, in contrast to lattice results \cite{Knechtli:2016pph}.
For instance, consider SU(3) Yang-Mills, cf.~Fig.~\ref{fig:YMeps}. The $\epsilon$ expansion in \cite{Morris:2004mg} yields for $\tilde{\alpha}=\frac{6}{(4\pi)^2}g^2$
 \bea
 \beta_{\tilde{\alpha}}&=& \epsilon\, \tilde{\alpha}- b_1\,  \tilde{\alpha}^2-b_2\,  \tilde{\alpha}^3-b_3\,  \tilde{\alpha}^4- b_4\,  \tilde{\alpha}^5,\\
 b_1&=& 3.67, \quad b_2=5.67, \quad b_3 =13.23, \quad b_4=39.43+\frac{51.22}{9},
 \eea
resulting in 
\be
\tilde{\alpha}_{\ast}=0.272 \epsilon - 0.115 \epsilon^2+0.024 \epsilon^3-0.016 \epsilon^4.
\ee
 
 Couplings which are marginally irrelevant in their critical dimension $d_c$ can achieve interacting fixed points for $d<d_c$, where they correspond to irrelevant directions. In contrast to the case in $d_c$, where the free fixed point results in a vanishing coupling at all scales in order to be a UV fixed point (triviality problem), in $d=d_c-\epsilon$, the interacting fixed point requires a unique finite value of the coupling in the IR, corresponding to the fixed-point value, unless the UV critical hypersurface is curved. Thus the interacting theory is UV complete for one unique value of the coupling. Conversely, asymptotically free trajectories reach the interacting fixed point in the IR.

 For instance, for scalar theories, the marginally irrelevant nature of the quartic coupling in $d=4$ implies the existence of a fixed point in $d=4-\epsilon$. The well-known Wilson-Fisher fixed point is IR attractive in the quartic coupling \cite{Wilson:1971dc} and serves as the IR endpoint of an asymptotically free trajectory. It has been characterized with various methods \cite{Guida:1998bx,Campostrini:1999at,Pelissetto:2000ek,Canet:2003qd,Litim:2010tt,ElShowk:2012ht,El-Showk:2014dwa,Gliozzi:2014jsa} and serves as a benchmark example for many techniques.
 For $d>4$, a possible fixed point \cite{Fei:2014yja} lies at negative quartic coupling, appearing to be at odds with a stable microscopic potential \cite{Percacci:2014tfa,Eichhorn:2016hdi}.

Fixed points generated by such a mechanism are weakly coupled at small $\epsilon$, where the critical exponent is equal to minus the canonical dimension. 

 A key example is gravity:
 Slightly above its critical dimension $d=2$, where the Einstein action is purely topological, the beta function for the dimensionless Newton coupling $G=G_N\, k^{d-2}$ at one loop reads \cite{Weinberg:1980gg,Gastmans:1977ad,Christensen:1978sc}, 
\be
\beta_G = \epsilon G - \beta_1\, G^2, \mbox{ such that } G_{\ast} = \frac{\epsilon}{\beta_1}, \quad \theta=- \epsilon,
\ee
where $\beta_1$ depends on the parameterization of metric fluctuations $h_{\mu\nu}$ around a background $\bar{g}_{\mu\nu}$. Note that in these calculations the Jacobian that arises in the path-integral measure from relating the different parameterizations is not taken into account.
Specifically the functional RG in the limit $d \rightarrow 2$ yields \cite{Percacci:2015wwa}
\bea
\beta_1&=&- \frac{2 \left( 19-38 \beta + 13 \beta^2\right)}{3(1-\beta^2)}, \quad \mbox{ for the linear parameterization, } g_{\mu\nu} = \bar{g}_{\mu\nu} + h_{\mu\nu}, \\
\beta_1&=& - \frac{2 \left(25-38 \beta + 19 \beta^2 \right)}{3(1- \beta)^2}, \quad \mbox{ for the exponential parameterization, } g_{\mu\nu} = \bar{g}_{\mu\kappa}{\rm exp[h..]}^{\kappa}_{\nu},
\eea
where $\beta$ is a gauge parameter, such that for $\beta \rightarrow 0$ the result $\beta_1=-38/3$ is found \cite{Tsao:1977tj,Brown:1976wc,Kawai:1989yh,Jack:1990ey} for the linear parameterization and $\beta_1=-50/3$ for the exponential parameterization \cite{Kawai:1992np,Kawai:1993mb,Kawai:1995ju,Aida:1994zc,Nishimura:1994qh,Aida:1996zn,David:1988hj,Distler:1988jt,Codello:2014wfa}. A continuous extension to $d=4$ might be possible \cite{Falls:2015qga,Falls:2017cze}.

\subsubsection{One-loop versus higher-loop}\label{sec:onevstwoloop}
\begin{figure}[!t]
\centering
\includegraphics[width=0.5\linewidth]{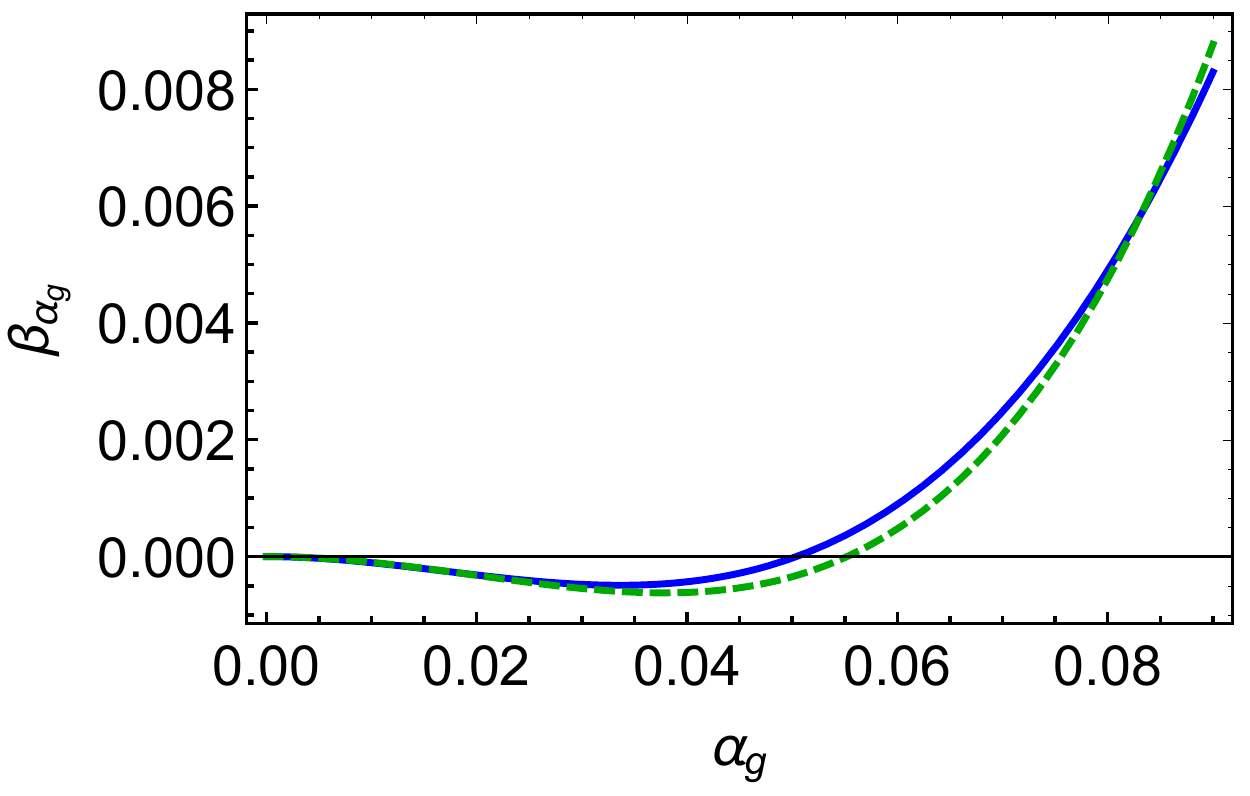}
\caption{\label{fig:betaLS}Beta function of the gauge coupling $\alpha_g$ for the model in \cite{Litim:2014uca} at next-to-leading order (two loop, blue continuous line) and next-to-next-to-leading order (three loop, green dashed line), cf.~Eq.~\eqref{eq:betaalphaLS} for the two-loop beta function. The fixed-point value for $\alpha_y$ at the corresponding order has been inserted and $\epsilon=0.1$ has been chosen. }
\end{figure}

In perturbation theory, the signs of the one-loop and two-loop coefficients can differ,  leading to a cancellation at a finite fixed-point value, schematically
\be
\beta_{g_i}=\beta_1\, g_i^{\#_1} + \beta_2\, g_i^{\#_2}+...
\ee
with 
\be
\beta_{g_i}\Big|_{g_i=g_{i\,\ast}}=0, \quad g_{i\,\ast}=\left(\frac{-\beta_1}{\beta_2} \right)^{\frac{1}{\#_2-\#_1}}.\label{eq:fp12loop}
\ee
The fixed point is real for ${\rm sign}(\beta_2)=-{\rm sign}(\beta_1)$. For it to lie at small values, where higher-loop terms are small, we require $|\beta_1|<|\beta_2|$. Actually,  the two-loop coefficient is a proxy for the higher-loop terms: the fixed point is generated by the competing signs of the one-loop versus the ``effective" two-loop term. As one extends the asymptotic perturbative series to higher loops, the fixed-point value shifts to compensate for the change, but as long as the sign of the ``effective" two-loop term is unchanged, a fixed point exists, cf.~Fig.~\ref{fig:betaLS}.

The interacting fixed point is IR attractive (repulsive) for $\beta_1>0\, (<0)$. Additionally, a UV (IR) attractive fixed point lies at $g_{i\, \ast}=0$.
Therefore, a complete trajectory exists between the free (interacting) fixed point in the UV and the interacting (free) fixed point in the IR for $\beta_{1}<0\, (>0)$. The former case is known as the Banks-Zaks fixed point in the case of non-Abelian gauge theories \cite{Caswell:1974gg,Banks:1981nn}. The latter underlies new developments in gauge-Yukawa models \cite{Litim:2014uca}, see Sec.~\ref{sec:GYAS}.

\subsubsection{Competing degrees of freedom}

In models with several degrees of freedom, a scale-invariant fixed-point regime can be achieved if the effect of different degrees of freedom can balance out - either within a perturbative expansion or at the nonperturbative level and for dimensionless as well as dimensionful couplings. 
Schematically,
\be
\beta_g = \beta_g^{({\rm d.o.f.}1)}-\beta_g^{({\rm d.o.f.}2)},
\ee
where, e.g., ${\rm d.o.f} 1$ might be a bosonic and ${\rm d.o.f.}2$ a fermionic contribution. 
($N=4$) super Yang Mills could be seen as a special example \cite{Sohnius:1985qm}.

A competition of fermionic and bosonic degrees of freedom also occurs in the beta function of a quartic scalar coupling which couples to fermions through a Yukawa coupling. This competition actually underlies Higgs mass bounds in the SM  \cite{Altarelli:1994rb,Hambye:1996wb}.
Perturbatively, the Yukawa coupling in simple Yukawa models is UV unsafe. Hints for a nonperturbative fixed point have been found \cite{Gies:2009hq,Gies:2009sv}, but been called into question in \cite{Vacca:2015nta} in extended truncations of the RG flow.

\section{Gauge-Yukawa models: Asymptotic safety at weak coupling}\label{sec:GYAS}

In $d=4$ dimensions, gauge-Yukawa models can exhibit perturbative asymptotic safety, discovered in \cite{Litim:2014uca}, achieved through a balance of one- versus two-loop effects. 
We follow \cite{Litim:2014uca} and consider a simple gauge theory with gauge coupling $g$ with
\be
\alpha_g = \frac{g^2}{(4\pi)^2}, 
\ee
with 2-loop beta function
\be
\beta_{\alpha_g} = \left(-B +C\, \alpha_g\right)\alpha_g^2.
\ee
An interacting fixed point lies at 
\be
\alpha_{g\, \ast}= \frac{B}{C}.
\ee
For the case $B>0, C>0$, this is the Banks-Zaks fixed point \cite{Banks:1981nn}, which is IR attractive in the gauge coupling. Accordingly, a complete RG trajectory can be constructed, emanating from the free fixed point in the UV and ending in a conformal regime in the IR. This can be achieved within the conformal window, e.g., $11/2 N_c < N_f< 34 N_c^3/(13 N_c^2-3)$ for $N_f$ fermions in the fundamental representation of $SU(N_c)$, \cite{Ryttov:2010iz,Pica:2010xq,Ryttov:2016ner}.\\
Asymptotic safety in the form of an IR-repulsive interacting fixed point occurs where asymptotic freedom is lost, i.e., the antiscreening effect of non-Abelian gauge bosons is overcompensated by the screening effect of charged matter. This requires $B<0$, see \cite{Caswell:1974gg, Tarasov:1976ef,Jones:1981we,Machacek:1983tz}, and accordingly $C<0$ for the coupling $g$ to be real. As shown in \cite{Caswell:1974gg}, see also \cite{Bond:2016dvk}, this is not possible to achieve with fermions only. Adding scalars to the model provides a Yukawa coupling 
\be
\alpha_y=\frac{y^2}{(4\pi^2)},
\ee
that results in
\be
C\rightarrow C - D\, \alpha_y.
\ee
This facilitates asymptotic safety. The one-loop Yukawa beta function reads
\be
\beta_{\alpha_y}= \partial_t\, \alpha_y = \left(E\, \alpha_y - F\, \alpha_g \right)\alpha_y,
\ee
see \cite{Fischler:1982du,Machacek:1983fi} for two-loop results.
The above system of beta functions admits three solutions
\bea
\alpha_{g\, \ast}&=&0, \, \alpha_{y\, \ast}=0,\\
\alpha_{g\, \ast}&=&\frac{B}{C}, \, \alpha_{y\, \ast}=0,\\
\alpha_{g\, \ast}&=&\frac{B}{C-D\frac{F}{E}}, \, \alpha_{y\, \ast}= \frac{B}{C-D\frac{F}{E}}\frac{F}{E},
\eea
where appropriate conditions on the coefficients of the beta function ensure that fixed-point values are real. The second fixed point is a generalization of the Banks-Zaks fixed point.
The fully interacting fixed point has one IR attractive and one IR-repulsive direction. The corresponding critical exponents are
\bea
\theta_{1,2}&=& \frac{-B\, E}{2(C\, E-D\, F)^2}\Bigl(-B\, C\, E - C\, E \, F + D\, F^2\nonumber\\
&{}& \pm \sqrt{B^2 C^2 E^2 - 2B\, F (C\, E-2D\, F)(C\, E- D\, F)+ F^2 (C\, E - D\, F)^2} \Bigr).
\eea
 Perturbative asymptotic safety in four-dimensional gauge theories requires the presence of fermions \emph{and} scalars \cite{Litim:2014uca}, providing a possible justification for the existence of fundamental scalars in nature. Moreover, (gravity-free) theories in four dimensions cannot exhibit weakly-coupled fixed points, i.e., arising from a balance of one-loop versus two-loop effects, unless gauge interactions are present \cite{Bond:2016dvk, Bond:2018oco}. This explains why tentative proposals for interacting fixed points in four-dimensional fermion-scalar theories lie in a nonperturbative regime \cite{Gies:2009hq,Eichhorn:2018vah}. 

As couplings can be rescaled arbitrarily (without an impact on the critical exponents), the fixed-point values of couplings do not automatically convey information on whether the fixed point is perturbative.
To achieve strict perturbative control over the fixed point, the critical exponents should be arbitrarily close to the canonical ones. This can be achieved in the Veneziano limit which allows to continuously emerge the fixed point from the free one.
 Hence we now focus on an $SU(N_c)$ gauge theory with $N_f$ flavours of Dirac fermions in the fundamental representation to take the Veneziano-limit,
\cite{Veneziano:1979ec},
\be
N_f\rightarrow \infty, \quad N_c\rightarrow \infty, \mbox{ with } \epsilon = \frac{N_f}{N_c}-\frac{11}{2}\, \mbox { finite}. \label{eq:Veneziano} 
\ee
$C=25$ \cite{Caswell:1974gg} holds in this limit without Yukawa interactions. 
The simplest way to add a $N_f\times N_f$ matrix $H$ of complex scalars is to have them uncharged under the gauge group,
\be
\mathcal{L}_{H-\rm pot}= - u \,{\rm Tr}\left( H^{\dagger}H\right)^2 - v\left({\rm Tr}H^{\dagger}H\right)^2.
\ee
 Then the two quartic  couplings decouple from the beta functions for the gauge and Yukawa coupling at the above order in perturbation theory and in the Veneziano limit \cite{Litim:2014uca}, see \cite{Jack:1983sk,Machacek:1984zw, Ford:1992pn} for the two-loop beta functions.

\begin{figure}[!t]
\includegraphics[width=0.45\linewidth]{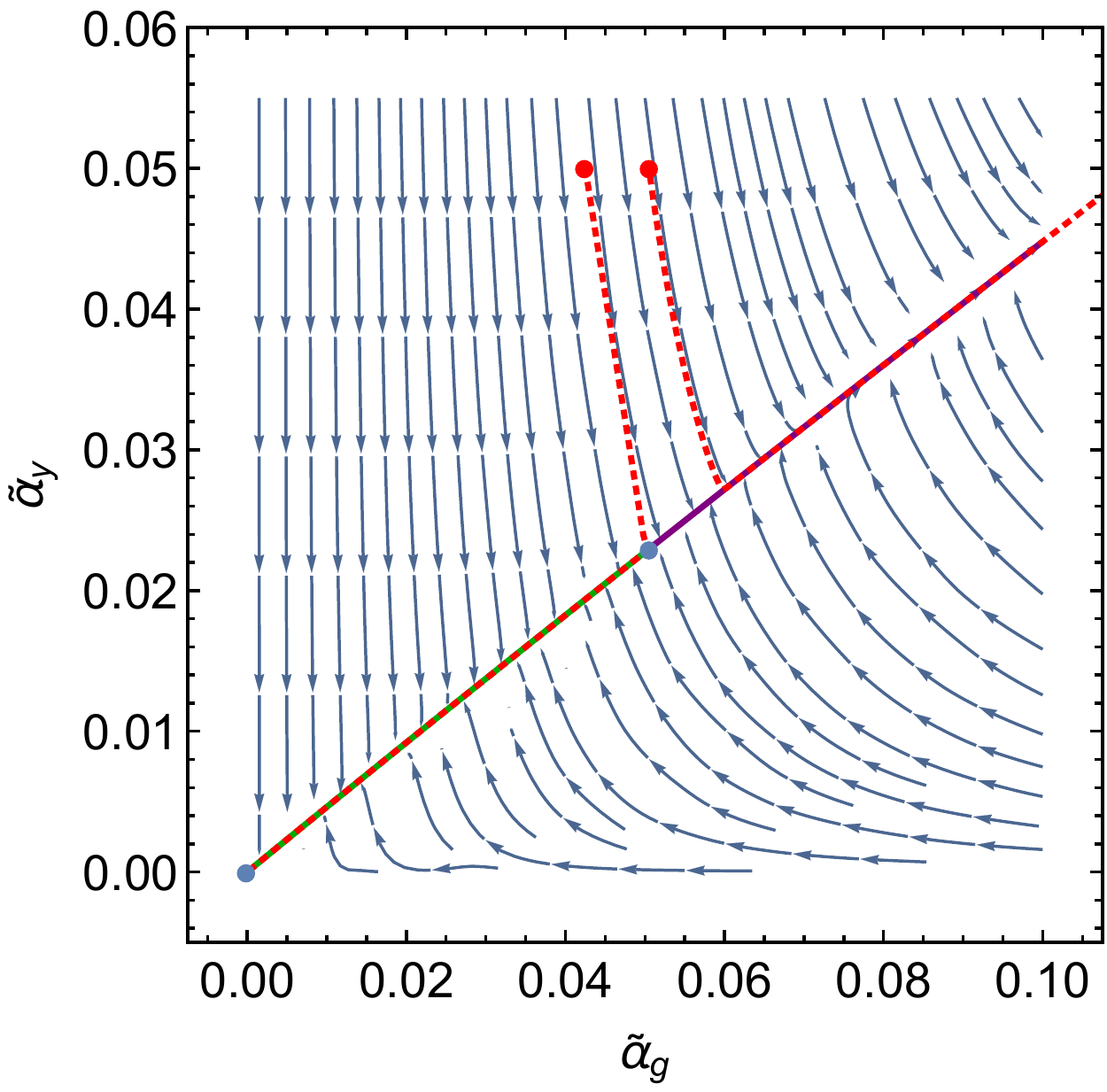}\quad \includegraphics[width=0.45\linewidth]{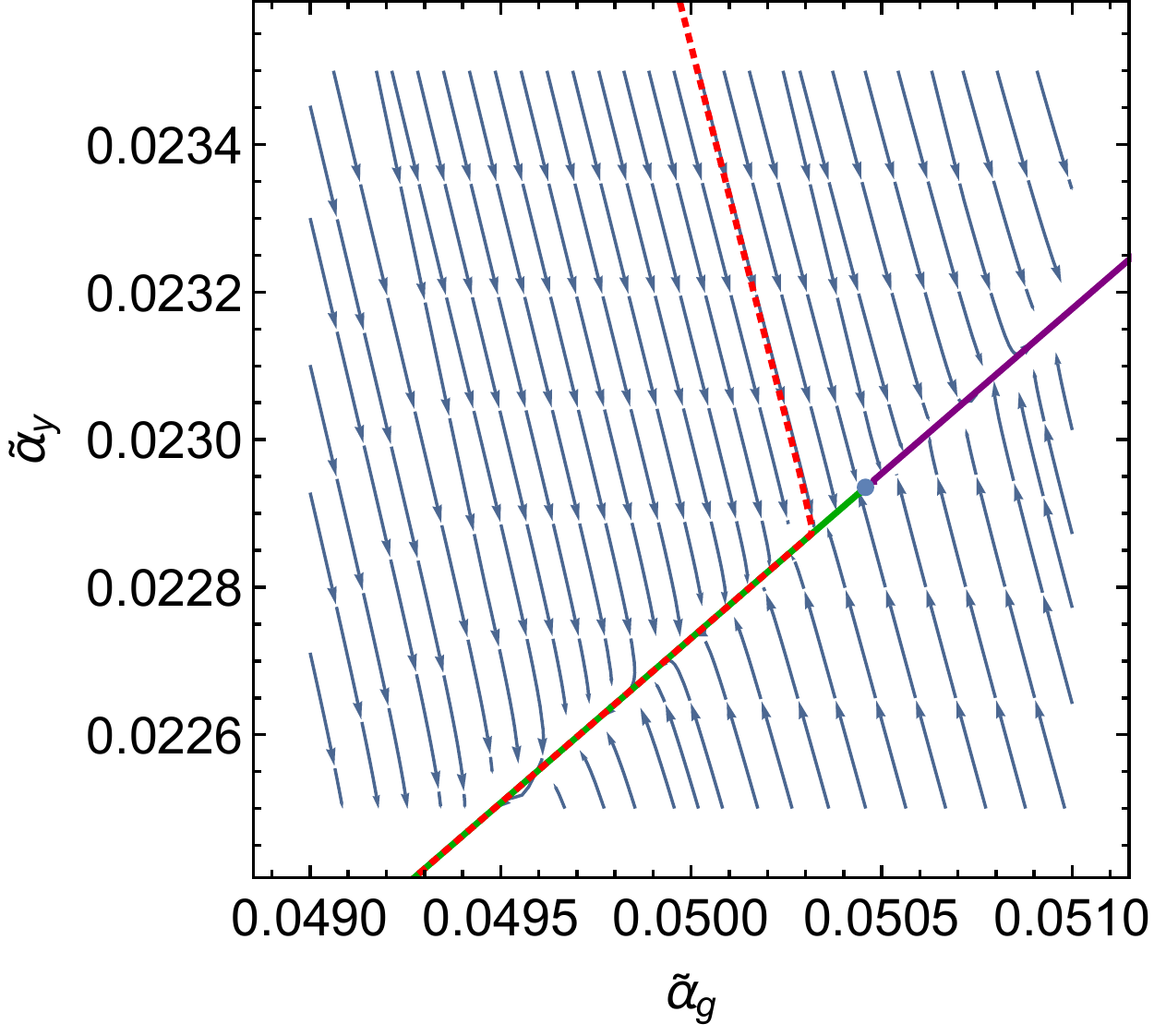}
\caption{\label{fig:LSflow} The flow towards the IR from the fixed point in Eq.~\eqref{eq:LSFP} for $\epsilon=1/10$ features one strongly IR attractive and one weakly IR repulsive direction. The green and purple, thick, continuous lines are the two only ``true" fixed-point trajectories. Initial conditions in the UV away from the fixed point (red dots) result in trajectories that are indistinguishably close to the fixed-point trajectories in the IR. The right panel shows a zoom into the vicinity of the fixed point, where the ``non-fundamental" trajectory narrowly misses the fixed point, but approaches the critical hypersurface arbitrarily closely, resulting in universal predictions in the IR.}
\end{figure}

In the limit \eqref{eq:Veneziano}, fixed-point values are controlled by $\epsilon$ and remain perturbative for $\epsilon <<1$ \cite{Litim:2014uca}. For a study of gauge groups and representations for which such a fixed point exists, see \cite{Bond:2017tbw}.
Asymptotic safety
is achieved in appropriately rescaled couplings that guarantee  well-behaved large-$N$-beta functions,
\be
\tilde{\alpha}_{y} =\frac{y^2\, N_c}{16\pi^2}, \quad \tilde{\alpha}_g = \frac{g^2\, N_c}{16\pi^2}.
\ee
The beta functions read
\bea
\beta_{\tilde{\alpha}_g} &=& \tilde{\alpha}_g^2 \left(\frac{4}{3}\epsilon + \left(25+\frac{26}{3}\epsilon \right)\tilde{\alpha}_g - 2 \left(\frac{11}{2}+\epsilon \right)^2 \tilde{\alpha}_y\right)\label{eq:betaalphaLS},\\
\beta_{\tilde{\alpha}_y}&=&\tilde{\alpha}_y \left( \left(13+2\epsilon \right)\tilde{\alpha}_y - 6 \tilde{\alpha}_g\right).
\eea
While the one-and two-loop contribution of the gauge coupling to $\beta_{\tilde{\alpha}_g}$ are positive, the contribution of the Yukawa coupling is negative. Accordingly a finite fixed-point value of the Yukawa coupling induces a physically acceptable fixed point, i.e., $\tilde{\alpha}_{g\, \ast}>0$. In turn, the positive contribution $\sim \tilde{\alpha}_y$ in $\beta_{\tilde{\alpha}_y}$ can balance against the negative one $\sim \tilde{\alpha}_g$ to generate a physically acceptable fixed point at $\tilde{\alpha}_y>0$. This results in an interacting fixed point emerging from the Gaussian one, since $\epsilon$ can become arbitrarily small for large enough numbers of fields,
\bea
\tilde{\alpha}_{g\, \ast}&=&\frac{26\epsilon+4 \epsilon^2}{57-46\epsilon-8 \epsilon^2},\nonumber\\
\tilde{\alpha}_{y\, \ast}&=&\frac{12\epsilon}{57-46\epsilon-8 \epsilon^2}.\label{eq:LSFP}
\eea
To leading order in $\epsilon$, the critical exponents are given by
\be
\theta_1=\frac{104}{171}\epsilon^2, \quad \theta_2 = -\frac{52}{19}\epsilon,
\ee
which go back to the canonical, vanishing values for $\epsilon\rightarrow 0$.
There is one IR repulsive and one IR attractive direction, fixing the Yukawa coupling at all scales in terms of the gauge coupling (or vice-versa). In other words, the value of one of the couplings in terms of the other is a prediction of the setting.\\
In a setting with ``non-fundamental" asymptotic safety (with new physics kicking in at some $k_{\rm UV}$), it is important that the velocity of the flow in the IR-attractive direction is of order $\epsilon$, whereas it is of order $\epsilon^2$ in the IR-repulsive direction. At the transition scale to the more fundamental description, initial conditions for the values of couplings are typically not the fixed-point values. Towards the IR, the flow is pulled towards the fixed point along the IR-attractive direction with a velocity $\mathcal{O}(\epsilon)$ and repelled from the fixed point along the IR-repulsive direction with a velocity $\mathcal{O}(\epsilon^2)$. Accordingly, near-fixed-point scaling could determine the behavior of a larger class of trajectories, cf.~Fig.~\ref{fig:LSflow}. 

At the next order in the approximation, quartic scalar self-interactions have to be included.
For the corresponding large $N$ couplings
\be
\alpha_h = \frac{u\, N_f}{16\pi^2}, \quad \alpha_v= \frac{v\, N_f}{16\pi^2},
\ee
the one-loop beta functions are given by \cite{Jack:1983sk,Machacek:1984zw, Ford:1992pn}
\bea
\beta_{\alpha_h}&=&-(11+2\epsilon)\tilde{\alpha}_y^2 + 4 \alpha_h \left(\tilde{\alpha}_y+2 \alpha_h \right),\label{eq:betaalphah}\\
\beta_{\alpha_v}&=& 12 \alpha_h^2+4 \alpha_v \left( \alpha_v+4 \alpha_h+ \tilde{\alpha}_y\right).
\eea
Due to the Yukawa coupling, fermionic fluctuations generate a scalar potential (cf.~first term in Eq.~\eqref{eq:betaalphah}) and cannot be set to zero consistently if $\tilde{\alpha}_{y\, \ast}\neq 0$; therefore a nontrivial fixed point of the system $\tilde{\alpha}_y, \tilde{\alpha}_g, \alpha_v, \alpha_h$ has to be found.
To satisfy Weyl consistency conditions (see below), the beta function of the gauge coupling is extended to three-loop order and that of the Yukawa coupling to two-loop order, where there is also a contribution $\sim \alpha_h$. The double-trace coupling $\alpha_v$ decouples from the remainder of the system.
The system admits a joint, asymptotically safe fixed point at $\alpha_{h\, \ast}>0$, and $\alpha_{v\, \ast}<0$ with $\alpha_{h\, \ast}+ \alpha_{v\, \ast}>0$, indicating a fixed-point potential that is bounded from below \cite{Litim:2014uca}. A study of the effective potential that includes quantum fluctuations at all scales on a trajectory emanating from the fixed point also indicates its stability \cite{Litim:2015iea}. At the fixed point, the scalar couplings are irrelevant, therefore the full model only features one free parameter.

The inclusion of two-loop effects in the gauge coupling and one-loop effects in the Yukawa coupling (or three-loop in the gauge, two-loop in the Yukawa, and one loop in the scalar couplings) is suggested by Weyl consistency conditions \cite{Jack:1990eb,Jack:2013sha}, which relate derivatives of beta functions. They arise by considering the model on a curved (but fixed) background and performing Weyl rescalings of the metric. As two  subsequent Weyl rescalings commute, it follows that  $\frac{\partial \beta^i}{\partial g_j} = \frac{\partial \beta^j}{\partial g_i}$. Herein $\beta^i=\chi^{ij}\beta_j$, where $\chi^{ij}$ is a metric in the space of couplings which depends on the couplings. An expression for $\chi^{ij}$ for gauge-Yukawa models has been derived in \cite{Antipin:2013pya}.

These conditions should hold for the full RG flow and can be imposed on the perturbative expansion. For a discussion of the corresponding ordering scheme for beta functions as well as other systematic choices of perturbative orders in the context of gauge-Yukawa theories, see also \cite{Bond:2017tbw}.

Residual interactions in canonically marginal couplings at an interacting fixed point provide finite contributions to beta functions of higher-order, canonically irrelevant couplings. Higher-order couplings in the scalar potential develop near-Gaussian fixed-point values of their own \cite{Buyukbese:2017ehm}. Accordingly, their scaling exponents follow the expectation that these couplings should remain irrelevant at a perturbative asymptotically safe fixed point.

The interacting fixed point in gauge-Yukawa systems constitutes a four-dimensional example of asymptotic safety, established within perturbation theory. It provides a new universality class which calls for an in-depth study of its possible extensions and generalizations.
  
The extension to a supersymmetric setting has been discussed in \cite{Intriligator:2015xxa,Bajc:2016efj,Bond:2017suy,Bajc:2017xwx}. While perturbative asymptotic safety cannot be realized in the supersymmetric setting with a simple gauge-group \cite{Intriligator:2015xxa}, it can exist in settings with semi-simple gauge groups \cite{Bond:2017suy,Bajc:2017xwx}. This highlights how an added symmetry can allow to derive strong no-go-theorems for asymptotic safety.

The fixed-point structure in gauge-Yukawa models is more intricate in a setting away from four dimensions \cite{Codello:2016muj} (or under the inclusion of potential quantum-gravity effects \cite{Christiansen:2017qca}), where the degeneracy of the free fixed point is lifted, and fixed-point collisions can occur.

Given that asymptotic safety appears in a range of gauge theories where asymptotic freedom is lost, the phase diagram of gauge theories could be richer than previously thought. In fact, indications for an interacting fixed point at leading order in $1/N_f$ go back to \cite{PalanquesMestre:1983zy,Gracey:1996he}, see \cite{Holdom:2010qs} for a recap and a discussion of higher orders in $1/N_f$.
With a view towards the potential phenomenological importance of such fixed points, \cite{Antipin:2017ebo, Antipin:2018zdg} employ a resummation of the fermionic bubble diagrams that contribute at leading order in $1/N_f$ to all orders in perturbation theory to the beta function for the non-Abelian gauge coupling. This provides indications for an interacting fixed point: In a $1/N_f$ expansion, the first nontrivial order vanishes in the large $N_f$ limit, unless there is a value of the coupling where it features a pole. In that case, depending on the sign of that contribution, a zero of the beta function can be generated. Indeed a corresponding pole can be found, providing an indication for a fixed point at a non-perturbatively large value of the gauge coupling.
A similar resummation for the gauge contribution to the leading nontrivial order of the beta function of the Yukawa coupling has been performed in \cite{Kowalska:2017pkt}.

The $a$-theorem \cite{Cardy:1988cwa} has been explored in this setting \cite{Dondi:2017civ,Antipin:2018brm}, showing that, as expected, the Jack and Osborn $a$ function \cite{Jack:2013sha} takes a larger value at the UV fixed point than at the IR fixed point. 

These developments pave the way for asymptotically safe model building beyond the Standard Model, e.g., \cite{Bond:2017wut,Abel:2017ujy,Mann:2017wzh,Abel:2017rwl,Molinaro:2018kjz}.

\subsection{Asymptotically safe phenomenology}
The idea that scale-invariance is realized in physics beyond the Standard Model has received a lot of attention, see, e.g., \cite{Meissner:2006zh,Shaposhnikov:2008xb,Shaposhnikov:2008xi,Khoze:2013uia,Holthausen:2013ota,Lindner:2014oea, Gies:2016kkk,Lewandowski:2017wov}, mostly focusing on settings with classical scale invariance. It is therefore highly intriguing to explore whether extensions of the Standard Model are asymptotically safe along the lines in \cite{Litim:2014uca}, realizing quantum scale invariance. Measurements showing a decreasing SU(3) coupling as a function of energy only cover a finite energy range and hence do not exclude asymptotic safety.

Steps towards an asymptotically safe Standard Model include the observation that
asymptotic safety can be achieved in semi-simple gauge groups \cite{Esbensen:2015cjw,Bond:2017lnq} and with chiral fermions \cite{Molgaard:2016bqf}. To render the non-Abelian Standard Model gauge couplings asymptotically safe, new fermionic states transforming in nontrivial representation of SU(2) and/or SU(3), have to be added.  Asymptotic safety might be achieved for one of the non-Abelian gauge couplings, with the others becoming asymptotically free, depending on the representation the new (vectorlike) fermions transform in \cite{Kowalska:2017fzw}. The matching scale, essentially corresponding to the mass scale of the new fermions, which separates the regime of power-law running below the fixed point from the regime of logarithmic running in the Standard Model, adds new free parameters to these models.\\
Yet, the non-Abelian gauge groups of the Standard Model are SU(2) and SU(3), not SU($N_c$) with $N_c\rightarrow \infty$, as required for the Veneziano limit. Accordingly, the addition of fermions to the Standard Model such that the non-Abelian gauge couplings, together with the BSM Yukawa coupling become asymptotically safe \cite{Bond:2017wut,Mann:2017wzh}, is difficult to reconcile with a perturbative nature of the extension \cite{Barducci:2018ysr}, at least if one also insists on solving the U(1) triviality problem. This is evident, e.g., in the large values of the critical exponents that lead to a fast flow away from the fixed point towards the IR, see, e.g., \cite{Bond:2017wut,Mann:2017wzh}. Hence, large $N_f$ fixed points \cite{PalanquesMestre:1983zy,Gracey:1996he,Holdom:2010qs,Antipin:2017ebo, Antipin:2018zdg,Kowalska:2017pkt} play a key role in these developments. Accommodating the Higgs at the correct mass is a challenge \cite{Pelaggi:2017abg}. This could change in a grand unified setting \cite{Molinaro:2018kjz}, which could also become asymptotically safe, \cite{Abel:2017rwl}.

Asymptotic safety beyond the Standard Model could have intriguing phenomenological consequences in astrophysics and cosmology \cite{Nielsen:2015una}. For instance, asymptotically safe dark matter could accommodate a significant running of the portal coupling to visible matter between the dark-matter-mass scale -- relevant for thermal production of dark matter in the early universe -- and the scale of direct detection experiments \cite{Sannino:2014lxa}. In the WIMP-paradigm, the dark matter relic density is linked to the probability of direct detection, since the cross-section for dark-matter-annihilation into Standard Model particles is related to the cross-section for dark matter scattering off Standard Model particles. Hence, the lack of direct detection has put severe constraints on the paradigm \cite{Tan:2016zwf,Akerib:2016vxi,Aprile:2018dbl}. These might be circumvented by introducing additional fields, providing a new parameter that decreases the tension between direct experimental bounds on the cross-section and the relic-density constraints.
Asymptotic safety could provide an alternative explanation \cite{Sannino:2014lxa}: the value of the coupling at the higher scale is larger as  it approaches an interacting fixed point. This might accommodate thermal production of  the full dark matter relic density  while being consistent with bounds  from direct searches.

\section{Asymptotically safe quantum gravity}
\subsection{Status of asymptotic safety in gravity}
Einstein gravity, quantized perturbatively, loses predictivity at (trans)planckian scales due to its perturbative nonrenormalizability. Infinitely many free parameters are associated with counterterms required to absorb new divergences appearing at every loop order \footnote{The enhanced symmetry in supergravity rules out many of these counterterms, shifting the expected order of divergence in the maximally supersymmetric theory to higher orders  \cite{Bern:2017ucb, Bern:2018jmv}.} \cite{'tHooft:1974bx,Deser:1974zza,Deser:1974zzc,Goroff:1985th,vandeVen:1991gw}. 
At momenta $p$ below the Planck scale, only a finite number of the counterterms contribute \cite{Donoghue:1993eb,Donoghue:1994dn} if one assumes that the corresponding dimensionless couplings are all of order one. Then, higher-order terms are suppressed by $(p/M_{\rm Planck})^\#$, $\#>2$.
Thus, gravity and quantum physics are actually compatible,  but a perturbative quantization only holds up to the Planck scale \cite{Donoghue:2012zc}. The key challenge is to find an ultraviolet completion. The minimalistic and conservative nature of asymptotic safety as compared to many other approaches to quantum gravity make it a useful starting point for this endeavour: If this ansatz for quantum gravity fails, more radical notions on the quantum nature of spacetime are required.

As the free fixed point is IR attractive in the Newton coupling, 
\footnote{In the higher-derivative theory with the additional invariants $R^2$ and $R_{\mu\nu}R^{\mu\nu}$  the marginal couplings are asymptotically free \cite{Stelle:1976gc,Fradkin:1981iu,Avramidi:1985ki}. Around flat space, this theory features a kinetic instability, see \cite{Salvio:2018crh} for a review. Breaking Lorentz symmetry allows to use higher-order spatial derivatives while keeping the action at second order in time derivatives \cite{Horava:2009uw}, resulting in  perturbatively renormalizability. Yet, the projectable version propagates an additional scalar that becomes nonperturbative in the IR. As the non-projectable version features a larger number of couplings, asymptotic freedom has only been established in $2+1$ \cite{Barvinsky:2017kob} dimensions and not $3+1$, as well as in the large $N$ limit for $N$ scalars coupled to Horava gravity \cite{DOdorico:2014tyh}. Constraints from pulsars \cite{Yagi:2013ava} and gravitational waves from a neutron-star merger \cite{Gumrukcuoglu:2017ijh} constrain these models.}
 the first mechanism for asymptotic safety cf.~Sec.~\ref{sec:canvsqm}, which is realized in $d=2+\epsilon$ dimensions  \cite{Weinberg:1980gg,Gastmans:1977ad,Christensen:1978sc,Tsao:1977tj,Brown:1976wc,Kawai:1989yh,Jack:1990ey,Kawai:1992np,Kawai:1993mb,Kawai:1995ju,Aida:1994zc,Nishimura:1994qh,Aida:1996zn,David:1988hj,Distler:1988jt,Codello:2014wfa}, might also determine the fate of gravity in $d=4$ dimensions. 
The physical mechanism behind asymptotic safety in gravity  \cite{Nink:2012vd} is the antiscreening nature of metric fluctuations that shield the Newton coupling, similar to the effect of self-interacting gluons in the Yang-Mills vacuum.

An extension of the $\epsilon$ expansion to higher order, combined with an appropriate resummation, could provide indications for or against a fixed point in four dimensions. This is also a goal of discrete approaches to the gravitational path-integral where spacetime configurations are constructed from scratch from microscopic building blocks:
Causal \cite{Ambjorn:2000dv,Ambjorn:2001cv} (and possibly also Euclidean  \cite{Laiho:2011ya,Laiho:2016nlp}) Dynamical Triangulations) (CDT) feature a higher-order phase transition \cite{Ambjorn:2011cg,Ambjorn:2012ij,Ambjorn:2017tnl} facilitating a continuum limit. This could provide a universality class for quantum gravity. 
Complementary to Monte Carlo simulations of dynamical triangulations, an analytical approach to search for a suitable continuum limit is based on tensor models \cite{Ambjorn:1990ge,Godfrey:1990dt,Gross:1991hx,Gurau:2016cjo,Benedetti:2011nn}, see Sec.~\ref{sec:RGQG}.
Lattice studies based on Euclidean Regge calculus have also been put forward as indications for asymptotic safety \cite{Hamber:2009mt,Hamber:2015jja}.\\
Intriguingly, perturbative techniques in $d=4$  yield indications for an asymptotically safe fixed point \cite{Codello:2006in,Niedermaier:2009zz,Niedermaier:2010zz}, providing a hint at a near-perturbative nature of asymptotically safe gravity.

\begin{figure}[!t]
\centering
\includegraphics[width=0.5\linewidth]{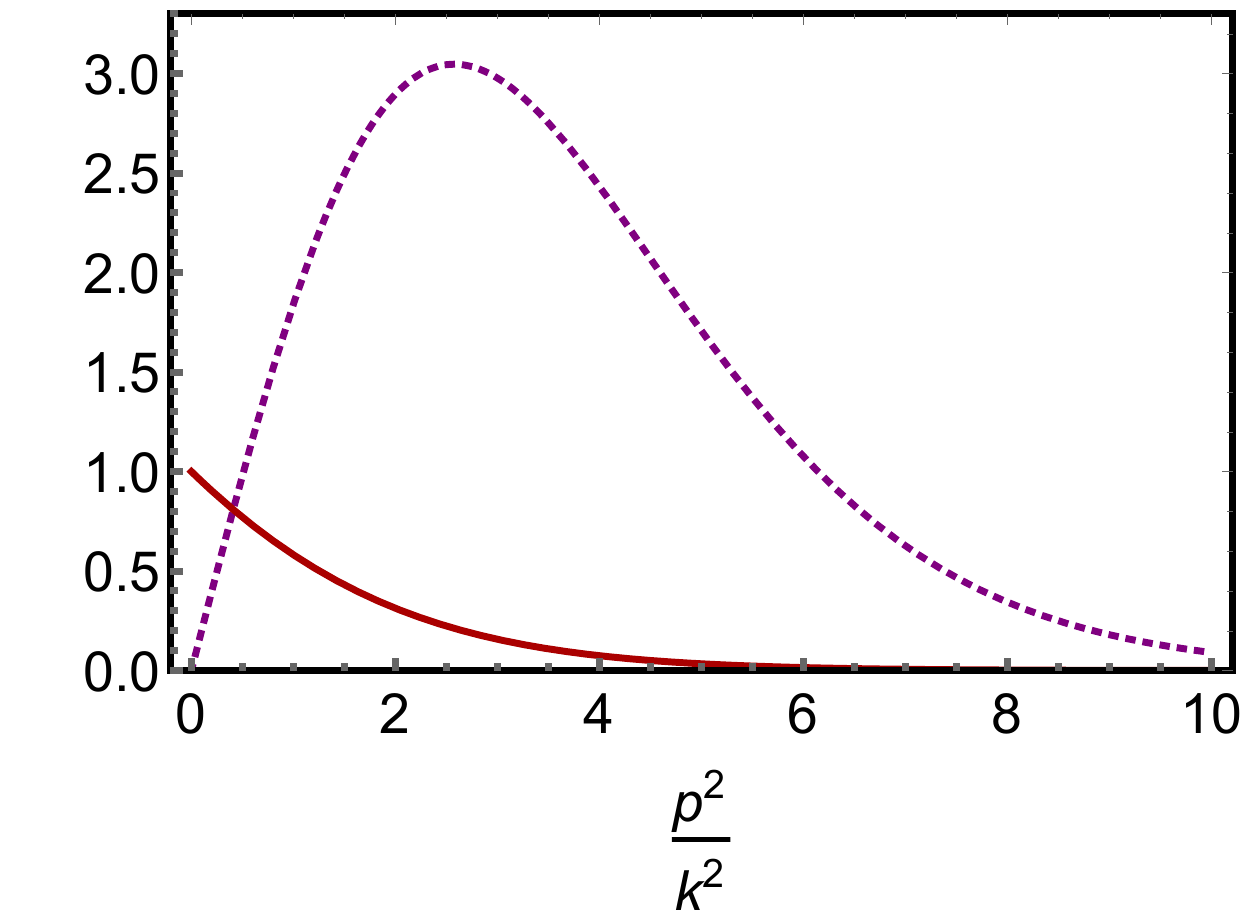}
\caption{\label{fig:reg}The regulator $R_k(p^2)$ (continuous red line) acts as a suppression term for IR modes. In the flow equation Eq.~\eqref{eq:floweq} its derivative with respect to $k$ (dotted purple line) acts as a suppression for UV modes, as well, such that the main contribution to the scale dependence of the dynamics at $k$ comes from modes at that momentum scale.}
\end{figure}

Most of the compelling evidence for asymptotic safety in gravity comes from Euclidean functional RG (FRG) studies based on the Wetterich equation
\footnote{A variant of the Polchinski equation also provides support for the asymptotic-safety conjecture \cite{deAlwis:2017ysy}.}. 
This framework provides beta functions for the dependence of couplings on the momentum scale $k$. The scale is introduced into the generating functional through an infrared cutoff function $R_k(p^2)$, called the regulator,
\bea
Z_k[J] &=& \int \mathcal{D}\varphi\, e^{-S[\varphi] - \frac{1}{2} {\rm Tr} \varphi R_k(p^2)\varphi + {\rm Tr}\int J\,\varphi},\\
\Gamma_k[\phi]&=& \underset{J}{\rm sup} \left( {\rm Tr}J \, \phi - {\rm ln}Z_k[J]\right)- \frac{1}{2} {\rm Tr} \phi R_k(p^2)\phi, \quad \quad \langle \varphi \rangle_k = \phi,
\eea 
reducing to the standard definitions at $k=0$. $R_k(p^2)$ and its derivative $k\partial_k\, R_k(p^2)$ are sketched in Fig.~\ref{fig:reg}.
This setup provides a flow equation, the Wetterich equation \cite{Wetterich:1992yh}, also
  \cite{Ellwanger:1993mw,Morris:1993qb}, pioneered for gauge theories in \cite{Reuter:1993kw} and gravity in \cite{Reuter:1996cp}. The regulator acts as a simultaneous IR- and UV cutoff, such that the change of a coupling at scale $k$ is mainly driven by quantum fluctuations at that scale:
\be
\partial_t =k\, \partial_k\Gamma_k = \frac{1}{2}{\rm Tr} \left[\left(\frac{\delta^2\Gamma_k[\phi]}{\delta\phi^2}+R_k \right)^{-1}\partial_t R_k\right].\label{eq:floweq}
\ee
For gravity, the covariant Laplacian $\bar{\Delta}$ with respect to an auxiliary background metric $\bar{g}_{\mu\nu}$ takes the role of the momentum $p^2$, \cite{Reuter:1996cp,Dou:1997fg,Souma:1999at,Lauscher:2001ya}.
 For general introductions and reviews, see \cite{Berges:2000ew, Polonyi:2001se,
Pawlowski:2005xe, Gies:2006wv, Delamotte:2007pf, Rosten:2010vm, Braun:2011pp}, for gravity, see \cite{Reuter:2012id,Nink:2013fkw,Ashtekar:2014kba,Percacci:2017fkn,Eichhorn:2017egq}. 
The method is  well-suited to models with dimensionful couplings, and therefore widely-used in condensed-matter physics and statistical physics \cite{Kopietz:2010zz,Metzner,Thomale}. Its relation to perturbation theory, which is straightforward at one loop, has been explored at higher loops in \cite{Papenbrock:1994kf,Codello:2013bra}. 

The FRG tracks the scale dependence of all couplings that are compatible with the symmetries, not just the perturbatively renormalizable interactions. For practical calculations, theory space is truncated to a (typically finite-dimensional) subspace, introducing  a systematic error. To highlight that quantitative results can already be achieved in relatively small truncations, we provide the leading scaling exponents for the Ising model in Tab.~\ref{tab:Ising}.

\begin{table}[!t]
\centering
\begin{tabular}{c|c|c|c}
truncation & $\nu=1/\theta_1$ & $\omega = - \theta_2$ & $\eta$\\\hline\hline
LPA 2 & 1/2 & 1/3 & 0\\
LPA 3 & 0.729 & 1.07 & 0\\
LPA 4 & 0.651 & 0.599 & 0\\
LPA 5 & 0.645 & 0.644 & 0\\
LPA 6 & 0.65 & 0.661 & 0\\
LPA 7 & 0.65 & 0.656 & 0 \\
LPA 8 & 0.65 & 0.654 & 0\\
\hline
LPA' 2 & 0.526 & 0.505 & 0.0546\\
LPA ' 3 & 0.684 & 1.33 & 0.0387 \\
LPA' 4 & 0.64 & 0.703 & 0.0433\\
LPA '5 & 0.634 & 0.719 & 0.0445\\
LPA' 6 & 0.637 & 0.728 & 0.0443\\
LPA ' 7 & 0.637 & 0.727 & 0.0443\\
LPA '8 & 0.637 & 0.726 & 0.0443.\\\hline\hline
\end{tabular}
\caption{\label{tab:Ising} 
Relevant and leading irrelevant critical exponent as well as the anomalous dimension for the Ising model obtained with the FRG in a derivative expansion to leading order (local potential approximation, LPA, to order $2n$ in the field) and next-to-order (LPA') with field-independent anomalous dimension. 
For the dimensionless potential $u[\rho] = \sum_{i=2}\frac{\lambda_i}{i!}(\rho-\lambda_1)^i$ with $\rho=\varphi^2/2$, the flow equation from which the beta functions for the couplings $\lambda_i$ are derived, reads
$\partial_t u[\rho]= -4 u+(d-2+\eta)\rho\, u'[\rho] + \frac{1}{2\cdot (4\pi)^2}\left(1-\frac{\eta}{6} \right)\frac{1}{1+u'[\rho]+2 \rho u''[\rho]}.$
The underlying derivation of the flow equation can be found, e.g., \cite{Berges:2000ew,Delamotte:2007pf} and the numerical evaluation of fixed-point values and critical exponents requires a basic numerical solver, such as Mathematica's FindRoot routine. At fourth order in the derivative expansion \cite{Canet:2003qd}, one obtains $\nu=0.632$ and $\eta=0.033$, see also \cite{Litim:2010tt} (compared to, e.g., $\nu=0.6304$ and $\eta=0.0335$ from 7-loop studies \cite{Guida:1998bx}).}
\end{table}
  
 For fixed points that arise via the mechanism in Sec.~\ref{sec:canvsqm}, the scaling is near-canonical near the critical dimension, providing a systematic way to devise truncations that include all relevant couplings. There are indications  that in quantum gravity four dimensions is close to two in the sense that the canonical dimension is a good predictor of relevance at the fixed point \cite{Falls:2013bv, Falls:2014tra}, enabling the setup of robust truncations by canonical power-counting. These indications require further confirmation, e.g., by including operators of the form $R \Box^n R$ \cite{deAlwis:2018azs}.

Considerable evidence for the existence of the interacting Reuter fixed point has accumulated, starting from the seminal work \cite{Reuter:1996cp} and \cite{Dou:1997fg,Souma:1999at,Lauscher:2001ya,Reuter:2001ag}, employing truncations of the form
\bea
\Gamma_k& &= -\frac{1}{16\pi G_N} \int d^4x\, \sqrt{g}\left(R- 2\Lambda \right) +\Gamma_{k\, \rm higher-order}\nonumber\\
&{}& + \frac{1}{32\pi G_N\, \alpha}\int d^4x \sqrt{\bar{g}}\bar{g}^{\mu\nu} \left(\bar{D}^{\kappa}h_{\mu\kappa} - \frac{1+\beta}{4}\bar{D}_{\mu}h \right)\left(\bar{D}^{\lambda}h_{\nu\lambda} - \frac{1+\beta}{4}\bar{D}_{\nu}h \right)\nonumber\\
&{}& - \sqrt{2}\int d^4x \sqrt{\bar{g}} \bar{c}_{\mu} \left(\left( \bar{g}^{\mu\rho}\bar{D}^{\kappa}g_{\rho\nu}D_{\kappa}+ \bar{D}^{\kappa}g_{\kappa\nu}D_{\rho}\right)- \frac{1+\beta}{2}\bar{D}^{\mu}D_{\nu}\right)c^{\nu}.
\eea
The third term is a gauge fixing term with two parameters $\alpha, \beta$ (see, e.g., \cite{Falkenberg:1996bq,Gies:2015tca,deBrito:2018jxt} for studies of the off-shell gauge dependence and \cite{Benedetti:2011ct} for gauge-independent on-shell results) and the third line is the corresponding Faddeev-Popov operator\footnote{A nontrivial wave-function renormalization \cite{Eichhorn:2010tb,Groh:2010ta} and ghost terms beyond the Faddeev-Popov are generated by the flow and have nonvanishing fixed-point values \cite{Eichhorn:2013ug}.}. Barred quantities refer to a background metric $\bar{g}_{\mu\nu}$ with respect to which the metric $g_{\mu\nu}$ can be gauge fixed, and a local coarse-graining scheme is set up. The fluctuation field is
\be
h_{\mu \nu}= g_{\mu\nu} - \bar{g}_{\mu\nu}.
\ee
A discussion of background-independence is given in Sec.~\ref{sec:bckrindep}. All results below are in the background-approximation, where $g_{\mu\nu} = \bar{g}_{\mu\nu}$ is used in the RG flow. Results from selected key truncations are summarized in Tab.~\ref{tab:FPgrav}. 

\begin{table}[!t]
\begin{tabular}{c|c|c|c|c|c|c|c|c}
ref. &gauge&cutoff& operators included & $\#$ rel. & $\#$ irrel.  & ${\rm Re}\theta_1$ & ${\rm Re}\theta_2$ & ${\rm Re}\theta_3$\\
&&& beyond  & dir. & dir. & & & \\ 
&&&Einstein-Hilbert & & & & & \\\hline\hline
\cite{Reuter:2001ag} & $\alpha=1, \beta=0$& exp.&- &2& - & 1.94 & 1.94 & -\\ \hline
\cite{Litim:2003vp} & $\alpha=0$&Litim\cite{Litim:2001up,Lauscher:2002sq}&- & 2& - & 1.67 & 1.67& - \\ \hline
\cite{Lauscher:2002sq}& $\alpha=0,\beta=0$& exp.& $\sqrt{g}R^2$ & 3 & 0 & 28.8 & 2.15 & 2.15 \\\hline
\cite{Machado:2007ea}&$\beta=1, \alpha=0$&Litim&$\sqrt{g}R^2,\sqrt{g}R^3$ & 3 & 1& 2.67& 2.67 &2.07  \\ \hline
\cite{Codello:2008vh}& $\alpha=1, \beta=1$& Litim & $\sqrt{g}R^2,\sqrt{g}R^3$ & 3 & 1 &2.71 &2.71 & 2.07\\\hline
\cite{Machado:2007ea}&$\beta=1, \alpha=0$&Litim&$\sqrt{g}R^2,\sqrt{g}R^6$ & 3 & 1& 2.39& 2.39 &1.51  \\ \hline
\cite{Codello:2008vh}& $\alpha=1, \beta=1$& Litim & $\sqrt{g}R^2,...,\sqrt{g}R^8$ & 3 & 6 &2.41 &2.41 & 1.40\\\hline
\cite{Falls:2013bv, Falls:2014tra} & $\alpha=0, \beta=0$ & Litim & $\sqrt{g}R^2,...,\sqrt{g}R^{34}$ & 3 &32 & 2.50 & 2.50 &1.59 \\ \hline
\cite{Benedetti:2009rx}&$\alpha=0$, h/o&Litim& $\sqrt{g}R^2,\, \sqrt{g}R_{\mu\nu}R^{\mu\nu}$ & 3 & 1 & 8.40 & 2.51 & 1.69 \\\hline
\cite{Gies:2016con}&$\beta=\alpha=1$ &Litim&$\sqrt{g}C^{\mu\nu\kappa \lambda}C_{\kappa \lambda \rho \sigma}C^{\rho \sigma}_{\,\,\,\,\,\,\mu \nu}$ & 2 & 1&1.48&1.48&- \\\hline
\end{tabular}
\caption{\label{tab:FPgrav} 
The operators beyond Einstein-Hilbert, the number of relevant/irrelevant directions, and the values of the positive critical exponents are indicated. All truncations listed above, employing the linear parameterization and single-metric approximation (cf.~Sec.~\ref{sec:bckrindep}) feature an asymptotically safe fixed point with no more than three relevant directions. (All results in the literature for finite-dimensional truncations feature an asymptotically safe form in qualitative agreement with these results.)}
\end{table}

For purposes of illustration, we also quote beta functions in the Einstein-Hilbert truncation with $G = G_N k^2$ and $\lambda = \Lambda/k^2$ from \cite{Codello:2008vh}, with anomalous dimensions $\eta_{h(c)}$ for the metric (ghost) (e.g., in \cite{Dona:2013qba}), and for the functional $f(\tilde{R})$, from \cite{Benedetti:2012dx} as found in \cite{Dietz:2012ic}.
\bea
\beta_G &=& 2 G - \frac{G^2}{12 \cdot 4 \pi}\left(\frac{52(4-\eta_h)}{1-2\lambda}+ 40(4-\eta_c) \right),\nonumber\\
\beta_{\lambda}&=&-2 \lambda +\frac{G}{12\cdot4\pi}\left(\frac{20(6-\eta_h)}{1-2\lambda}-16(6-\eta_c) \right) - \frac{G\, \lambda}{12\cdot 4\pi}\left(\frac{52 (4-\eta_h)}{1-2\lambda}+40 (4-\eta_c) \right),\\
\partial_t f&=&4f-2\tilde{R}\, f' +\frac{1}{384\pi^2}\Bigl[-20\frac{\partial_t f'-2\tilde{R}\, f''+8f'}{(\tilde{R}-2)f'-2f} -36 -12-5\tilde{R}^2 \label{eq:flowfR}\\
&{}&+ \frac{(\tilde{R}^4-54\tilde{R}^2-54)(\partial_t f''-2\tilde{R}\,f''')-(\tilde{R}^3+18 \tilde{R}^2+12)(\partial_tf'-2\tilde{R}\,f''+2f')-36(\tilde{R}^2+2)(f'+6f'')}{2\left(-9f''+(\tilde{R}-3)f'-2f \right)}\Bigr].\nonumber
\eea
Here, $\tilde{R}=R/k^2$ is the dimensionless curvature and primes denote derivatives with respect to $\tilde{R}$. Eq.~\eqref{eq:flowfR} provides the beta functions for couplings of $R^n$ upon a Taylor expansion in the curvature.

\begin{figure}[!t]
\includegraphics[width=0.45\linewidth]{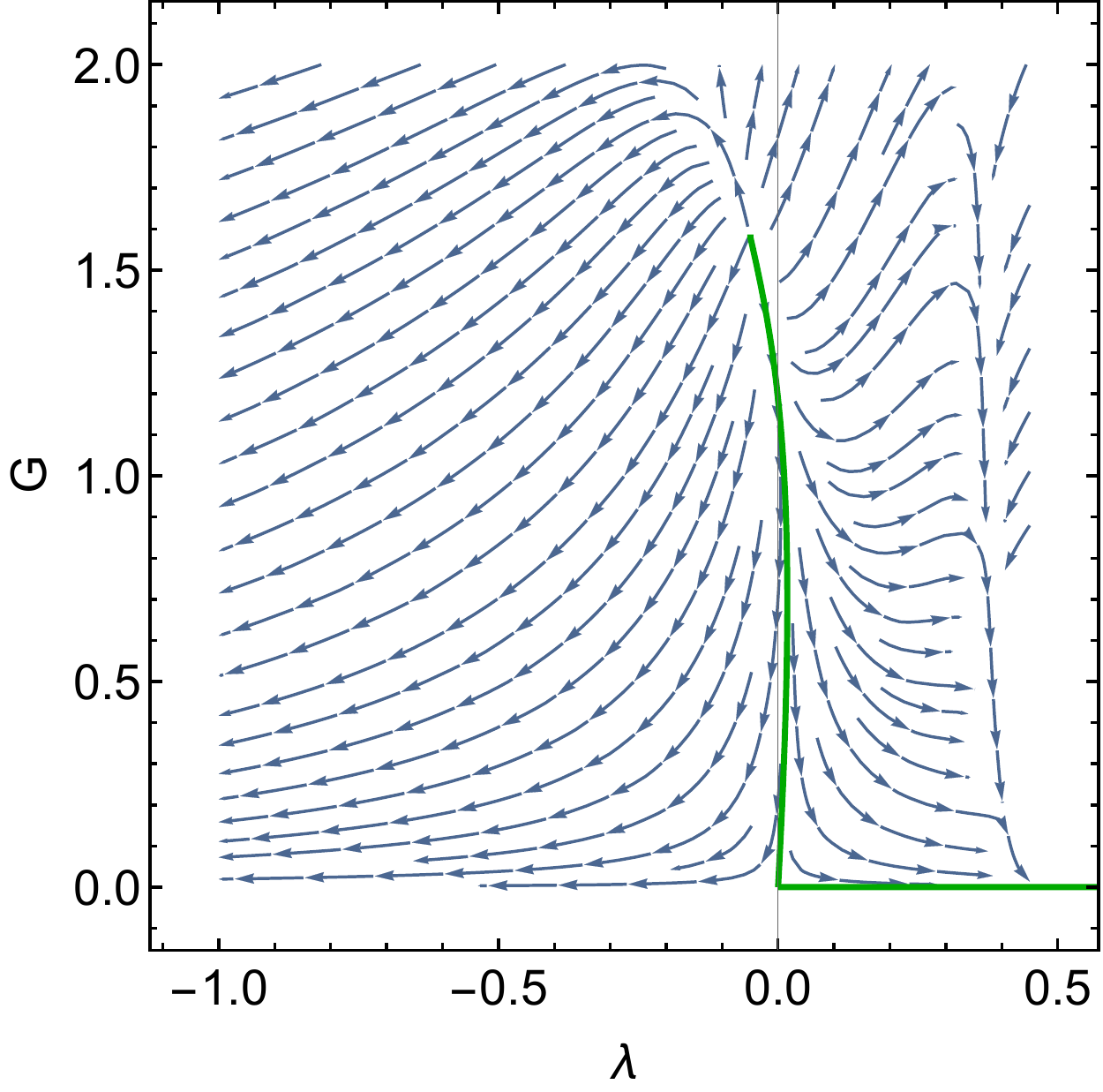} \quad \includegraphics[width=0.45\linewidth]{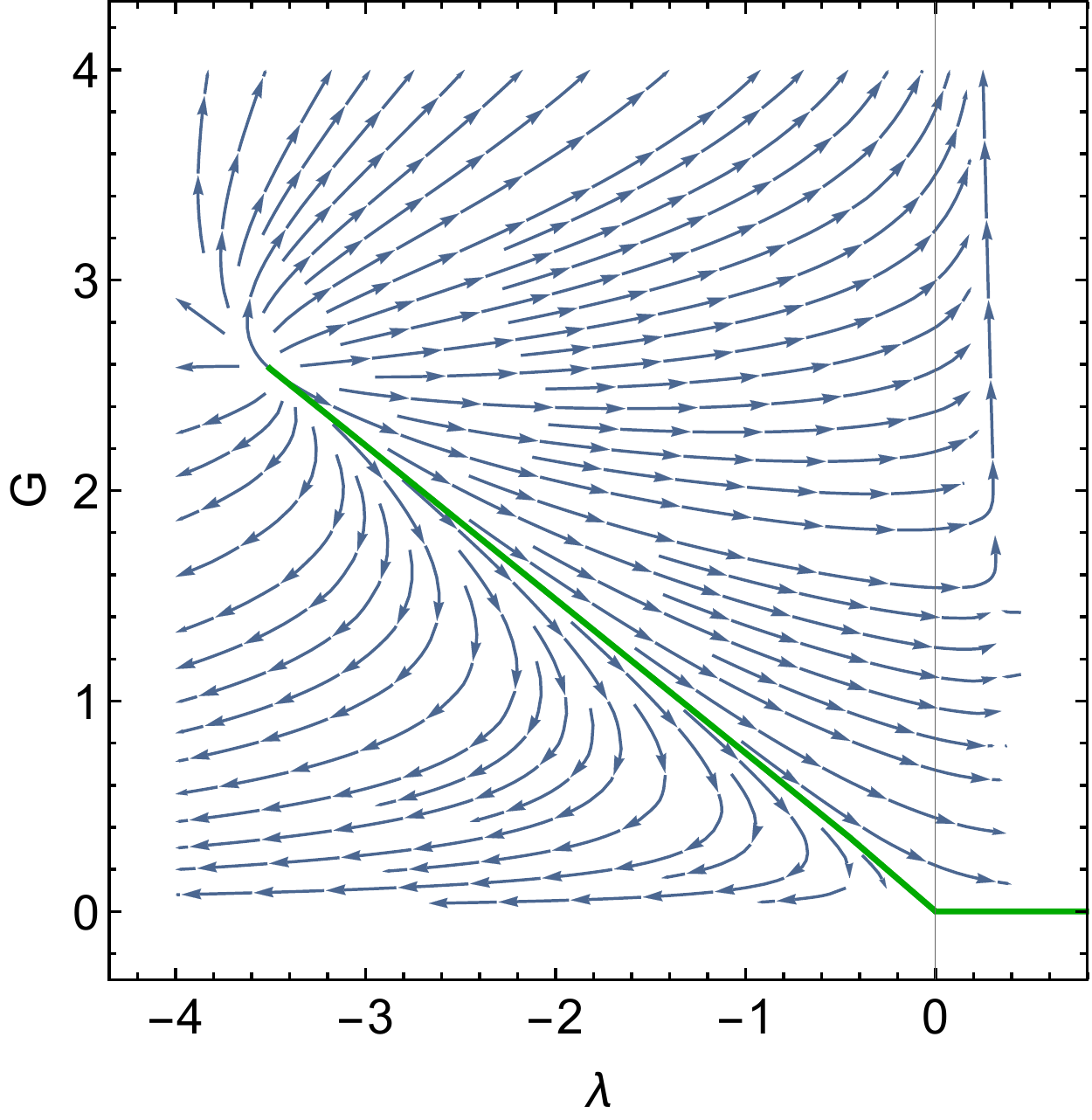}
\caption{\label{fig:RGFlowgravity} The RG flow to the IR in the Einstein-Hilbert truncation in the setup discussed in \cite{Dona:2013qba} for a type Ia cutoff features a trajectory -- passing very close to the free fixed point --  on which the dimensionful Newton coupling and cosmological constant reach constant values in the IR in agreement with measurements. Left panel: pure gravity case; right panel: including minimally coupled matter as in the Standard Model (4 scalars, 12 vectors, 45 Weyl fermions). The two eigendirections of the fixed point are superposition of $G$ and $\lambda$.}
\end{figure}

The Newton coupling and cosmological constant are relevant, cf.~Fig.~\ref{fig:RGFlowgravity}. Accordingly, the IR value of the cosmological constant is unrestricted. The choice of different fixed-point trajectories in Fig.~\ref{fig:RGFlowgravity} results in different values of the dimensionful cosmological constant in the IR $\Lambda_{\rm IR}$. To realize $\Lambda_{\rm IR}/M_{\rm Pl}^2<<1$, a specific trajectory has to be chosen. The question, why this particular trajectory is realized, is the finetuning ``problem". Yet, any relevant coupling is actually linked to a similar question. For instance, the value of the QCD coupling at the electroweak scale would be different on other, also asymptotically free trajectories. The question how a more fundamental principle selects one out of many viable trajectories for the relevant couplings exists irrespective of whether the coupling is logarithmically or power-law sensitive to the momentum scale. (The need for successive tuning at each order in perturbation theory for the power-law case is a consequence of that particular approximation scheme, not a signature of a consistency problem of the theory. )

\begin{figure}[!t]
\includegraphics[width=0.45\linewidth]{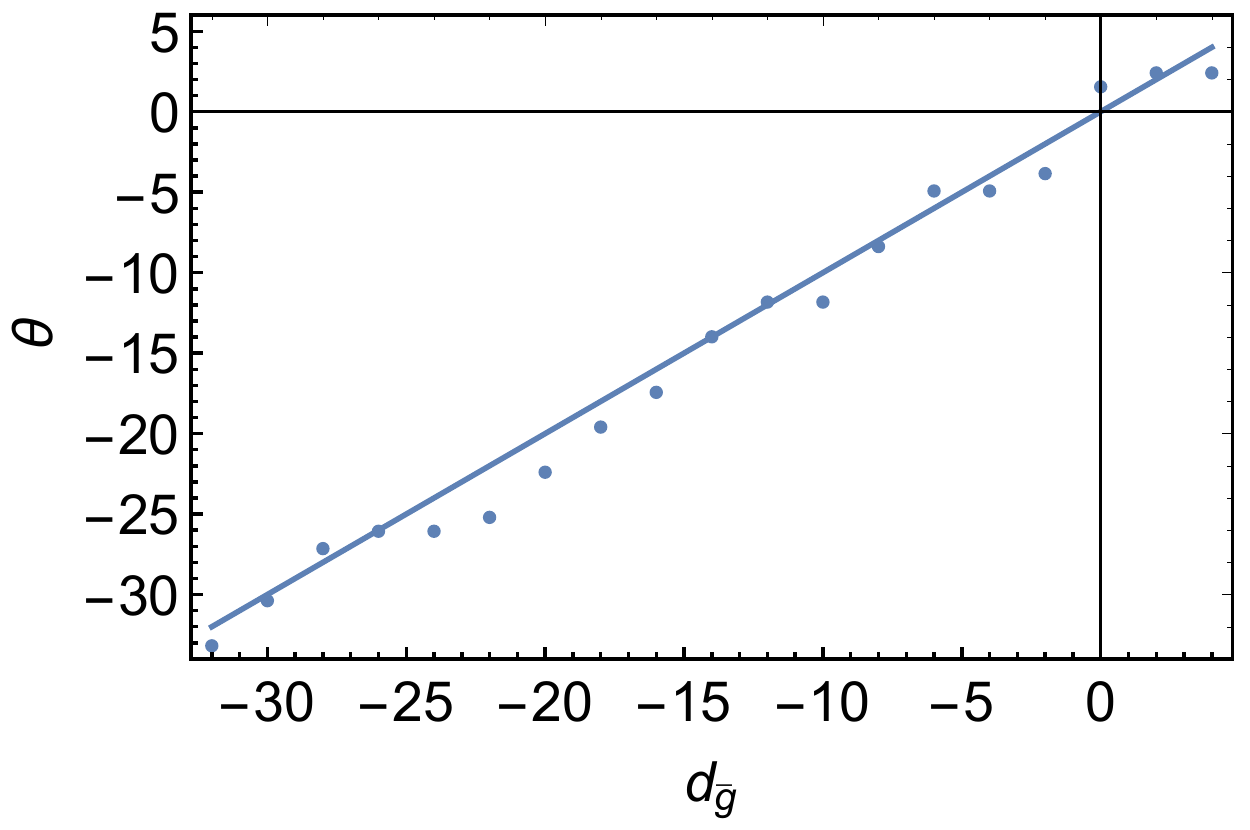}\quad \includegraphics[width=0.45\linewidth]{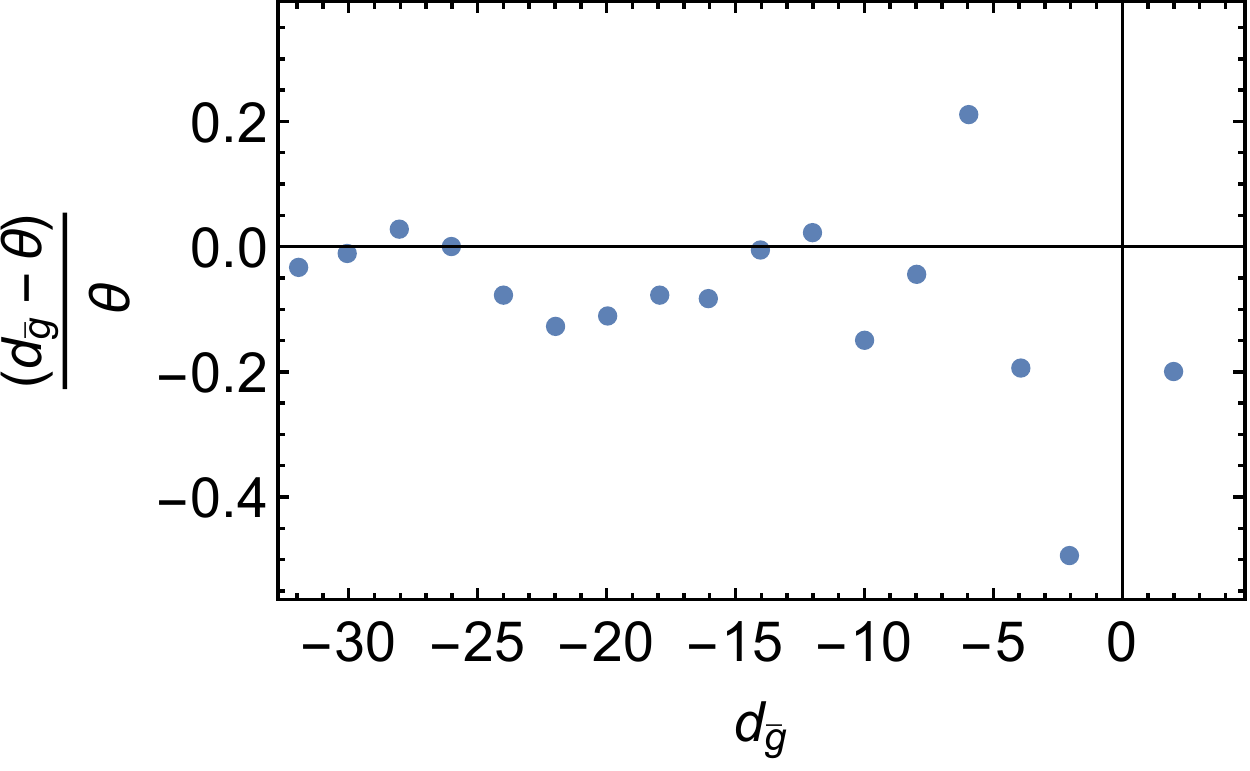}
\caption{\label{fig:GaussRn} Data from \cite{Falls:2014tra} on the critical exponents in a truncation $\sum_{n}\sqrt{g}R^n$, compared to the canonical dimension.}
\end{figure}

At the interacting Reuter fixed point, canonical ordering appears to hold \footnote{A combined truncation of \cite{Gies:2016con} with the operators in \cite{Benedetti:2009rx} remains to be explored.}, cf.~Tab.~\ref{tab:FPgrav}.
This provides a scheme to set up consistent truncations:
Assume for simplicity that the operators in Tab.~\ref{tab:FPgrav} diagonalize the stability matrix, resulting in critical exponents
\be
\theta_i = d_{\bar{g}_i}+ \eta_i.
\ee
Unless the anomalous scaling contribution $\eta_i$ would grow with the canonical dimension, the canonical dimension dominates for canonically highly irrelevant couplings, rendering them irrelevant at an interacting fixed point.
In fact, already at the level of canonically marginal couplings of $R^2$ and $R_{\mu\nu}R^{\mu\nu}$, there appears to be only one relevant direction\footnote{Results in the exponential parameterization even yield one relevant direction less \cite{Ohta:2015fcu,deBrito:2018jxt}.}.
All canonically irrelevant operators that have been examined are irrelevant at the fixed point. In \cite{Falls:2013bv, Falls:2014tra}, the normalized difference of canonical and quantum scaling dimension decreases with decreasing canonical dimension for $R^n$, cf.~Fig.~\ref{fig:GaussRn}. The near-Gaussian scaling spectrum (at higher orders in the curvature expansion) is also in line with the possibility to find indications of asymptotic safety from perturbation theory \cite{Niedermaier:2009zz,Niedermaier:2010zz}.

Systematic truncation errors can be estimated given that in approximations schemes for QFTS, dependencies on unphysical parameters arise even at the level of observables.  The better the approximation, the weaker such a dependence. Tests include gauge-parameter dependence \cite{Gies:2015tca}, regulator dependence \cite{Reuter:2001ag} and dependence on the parameterization for metric fluctuations \cite{Gies:2015tca,Ohta:2016npm,deBrito:2018jxt}. 

Going beyond finite-dimensional truncations, the closed fixed-point equation for $f(\tilde{R})$, e.g., Eq.~\eqref{eq:flowfR} has been investigated. Depending on the choice of regulator, it contains a varying number of fixed singularities, as the regulator introduces additional field-dependence in the background approximation.  
Thus, specific choices of the regulator allow for global solutions \cite{Benedetti:2012dx,Demmel:2012ub,Dietz:2012ic,Dietz:2013sba,Demmel:2015oqa,Ohta:2015efa} while others do not \cite{Codello:2008vh}. One might conclude that extensions of the truncation are required, going beyond the background- approximation for $f(\tilde{R})$, see \cite{Christiansen:2017bsy}, see Sec.~\ref{sec:ASquestions}.

Many  gravitational theories are classically dynamically equivalent to GR. Thus different theory spaces could allow for asymptotic safety \cite{Krasnov:2017epi}. For instance, the vielbein and the connection can be treated as independent variables \cite{Daum:2010qt,Daum:2013fu,Harst:2014vca,Harst:2015eha}, or torsion can be included \cite{Pagani:2015ema,Reuter:2015rta}. 
The dimension of theory space and the number of relevant couplings decrease by one \cite{Eichhorn:2015bna} in unimodular gravity \cite{Unruh:1988in,Finkelstein:2000pg,Ellis:2010uc}, where the determinant of the metric is a fixed density,  removing the cosmological constant from the action. 
Further, fluctuations in topology, dimensionality, signature etc. might be included in the gravitational path integral. The corresponding additional configurations either prevent the existence of a continuum limit/ RG fixed point, lead to an asymptotically safe fixed point in the same universality class as Tab.~\ref{tab:FPgrav}, or provide another gravitational universality class which differs in its physical implications and can therefore (in principle) be probed experimentally.

In two dimensions, the conformal field theory underlying asymptotic safety has been studied \cite{Nink:2015lmq}. In $d=4$, scale-invariance need not imply conformal invariance (in fact, sufficient conditions for this are not known). If it were possible to extend the conformal bootstrap program \cite{Simmons-Duffin:2016gjk} to a gravitational setting, a search for the corresponding universality class with relevant directions according to Tab.~\ref{tab:FPgrav} might answer whether there is a conformal theory behind asymptotically safe gravity.

\subsection{Open questions \& future perspectives}\label{sec:ASquestions}
\subsubsection{Lorentzian signature}
There is no simple Wick-rotation in quantum gravity, thus the above results do not directly imply Lorentzian asymptotic safety. 
In an ADM decomposition of the metric, the change of signature can be implemented by changing one parameter. This has been used in \cite{Manrique:2011jc} to find hints for asymptotic safety in a Lorentzian setting for the Einstein-Hilbert truncation. Further, RG flows in the ADM decomposition have been explored in \cite{Rechenberger:2012dt,Biemans:2016rvp,Biemans:2017zca}.  The  FRG can be formulated in a Lorentzian setting \cite{Floerchinger:2011sc}, underlying the study of real-time correlators, e.g., in QCD \cite{Pawlowski:2015mia}. 

Alternatively, a proposal \cite{Eichhorn:2017bwe} to search for Lorentzian asymptotic safety employs causal set quantum gravity. This is an intrinsically Lorentzian, discrete approach to quantum gravity, based on the path integral over all causal sets \cite{Bombelli:1987aa}. Under the restriction to manifoldlike causal sets (implemented as a path integral over sprinklings \cite{Henson:2006kf,Dowker:aza}) the space of couplings might feature a second-order phase transition \cite{Surya:2011du,Glaser:2018jss}.

\subsubsection{Propagating degrees of freedom}
Higher-order derivatives in QFTs on a flat background generically imply an instability in the kinetic term \cite{Ostrogradsky:1850fid,Woodard:2015zca}, translating into a violation of reflection positivity for the Euclidean propagator \cite{Arici:2017whq}. In a quantum setting, the unboundedness of the Hamiltonian can be traded for unitarity violation through negative-norm states in the Hilbert space \cite{Woodard:2015zca}. \\
In quantum gravity, an analysis of unitarity  is presumably rather more subtle for several reasons.\\ 
Firstly, positivity violation in gauge-variant propagators occurs in unitary theories such as QCD \cite{Cucchieri:2004mf,Bowman:2007du}. (A direct analogy with QCD has been proposed for (asymptotically free higher-derivative) gravity in \cite{Holdom:2015kbf,Holdom:2016xfn}.)
The physical ``graviton" as the transverse traceless part of the metric propagator is defined perturbatively; but non-perturbatively no local separation of gauge and physical degrees of freedom is possible.\\
Secondly, an instability in the flat-space propagator is not in conflict with observations, given that the cosmological background appears to be FRW-like.\\
Thirdly, Ostrogradski instabilities occur under a crucial assumption, namely that of finitely many higher-order terms. Yet the case with infinitely higher order terms \emph{can} feature a well-defined propagator, translating into a well-posed initial value problem at the level of the equations of motion \cite{Barnaby:2007ve}. Examples include string-field theory, see \cite{Barnaby:2007ve} and references therein. Accordingly, truncated dynamics in asymptotically safe gravity might contain spurious instabilities (just as an analysis of a truncated effective action for string theory would).\\
Fourth, even at the level of curvature-squared actions, the mass of the ``ghost" (analyzed around \emph{flat} space) runs as a function of momentum. Hence, such ghosts might not appear as physical states, see  \cite{Floreanini:1994yp,Benedetti:2009rx,Becker:2017tcx}. \\
Finally, if asymptotic safety is ``non-fundamental" (cf.~Sec.~\ref{sec:ASnonfund}), the mass-scale of the ghosts (if these exist on physically relevant backgrounds) sets an upper bound on $k_{\rm UV}$.

CDT satisfies reflection positivity \cite{Ambjorn:2000dv,Ambjorn:2001cv}. Thus its continuum limit, which might correspond to asymptotically safe gravity, inherits this property. As many other examples, this reinforces that the quest to understand quantum spacetime can be accelerated by searching for links between quantum-gravity approaches.

In addition to ghost-like states, higher-order gravity can (but again, need not) contain additional propagating degrees of freedom. These might be of phenomenological interest, e.g., driving inflation or leading to modifications of GR detectable in black holes and/or gravitational waves.

Determining the spectrum of propagating gravitational degrees of freedom in asymptotically safe gravity is an important outstanding question. A comprehensive answer in the FRG approach requires studying the full propagator (at $k=0$, where all quantum fluctuations contribute) around a solution to the quantum equations of motion.

\subsubsection{Background independence}\label{sec:bckrindep}

Background independence is a key property of quantum gravity, meaning that all configurations in the path integral should be treated on an equal footing. This appears to be at odds with the introduction of a local coarse graining scheme, as this relies on a metric. Specifically, the regulator in the flow equation depends on a background metric $\bar{g}_{\mu\nu}$. 
Additionally, a local formulation of gauge theories requires gauge fixing to derive the propagator. The flow equation is based on a background gauge-fixing. 
Nevertheless, background independence can be achieved, if all backgrounds are treated on the same footing \cite{Becker:2014qya}, i.e., if $g_{\mu\nu}$ and $\bar{g}_{\mu\nu}$ are both kept as distinct arguments of the flowing action. In the limit $k \rightarrow 0$, where the regulator vanishes, setting $g_{\mu\nu}= \bar{g}_{\mu\nu}$ yields an effective action that inherits diffeomorphism invariance and therefore background independence from the auxiliary background-diffeomorphism invariance that is kept intact for an appropriate choice of gauge fixing and regulator function.  Therefore, ultimately we are interested in  $\Gamma_{k\rightarrow 0}[\bar{g}_{\mu\nu}, g_{\mu\nu}= \bar{g}_{\mu\nu}]$, or $\Gamma_{k\rightarrow 0}[\bar{g}_{\mu\nu}, h_{\mu\nu}=0]$, respectively. Crucially, the flow is driven by the fluctuation propagator, $\left(\Gamma_k^{(0,2)}[\bar{g}_{\mu\nu}, g_{\mu\nu}]\right)^{-1} = \left(\frac{\delta^2}{\delta g_{\kappa \lambda}\delta g_{\rho\sigma}}\Gamma_k[\bar{g}_{\mu\nu},g_{\mu\nu}] \right)^{-1}$, or, equivalently, $\left(\Gamma_k^{(0,2)}[\bar{g}_{\mu\nu}, h_{\mu\nu}]\right)^{-1}$. As the regulator and gauge fixing break the symmetry between $g_{\mu \nu}$ and $\bar{g}_{\mu\nu}$, this is not the same as $\left(\Gamma_k^{(2,0)}[\bar{g}_{\mu\nu}, g_{\mu\nu}]\right)^{-1}$. 
 Schematically,
\be
\partial_t \Gamma_k[\Phi_{\rm phys}, \Phi_{\rm bck}]= \frac{1}{2}{\rm Tr}\left[\left(\frac{\delta^2\Gamma_k[\Phi_{\rm phys}, \Phi_{\rm bck}]}{\delta \Phi_{\rm phys}^2}+R_k[\Phi_{\rm bck}] \right)^{-1}\partial_t R_k[\Phi_{\rm bck}]\right].
\ee
In the background approximation, one equates $\Phi_{\rm phys} = \Phi_{\rm bck}$ after the derivation of $\frac{\delta^2\Gamma_k[\Phi_{\rm phys}, \Phi_{\rm bck}]}{\delta \Phi_{\rm phys}^2}$. Accordingly, projections on field monomials pick up the auxiliary background-field dependence of the regulator in this approximation.

As an intermediate step to obtaining an effective action that respects background independence, one  has to derive the flow of the fluctuation field propagator \cite{Christiansen:2012rx,Christiansen:2014raa,Christiansen:2015rva,Denz:2016qks} in a setting that makes explicit use of a background. Alternatively, one can map this to a ``bimetric" truncation, in which the propagator of the full metric is distinguished from the background metric, and drives the RG flow \cite{Manrique:2009uh,Manrique:2010mq,Manrique:2010am,Becker:2014qya}, see Tab.~\ref{tab:bimetric}.

\begin{table}[!t]
\centering
\begin{tabular}{c|c|c|c|c|c|c|c}
$G_{\ast}$&$\lambda_{\ast}$ & $G_{B\,\ast}$ &$\lambda_{B\, \ast}$& $\theta_1$ & $\theta_2$ & $\theta_{B\, 1}$& $\theta_{B\,2}$\\\hline\hline
0.70 & 0.21 &8.2 &-0.01 & 3.6+4.3i & 3.6 -4.3i & 4&2
\end{tabular}
\caption{\label{tab:bimetric}Fixed-point results from \cite{Becker:2014qya} for the ``dynamical" couplings in the Einstein-Hilbert truncation and their background counterparts. Critical exponents can be split into the two sectors, as the background couplings do not couple into the flow of the dynamical couplings and accordingly the stability matrix is upper/lower triangular in the background sector yielding canonical exponents.}
\end{table}

The fluctuation-field dynamics are not protected by an auxiliary diffeomorphism invariance (as the background dynamics is). Accordingly, the theory space is that of a spin-2-field, with (modified) Slavnov-Taylor identities relating different couplings as a consequence of the symmetry. In a vertex expansion, this results in distinct ``avatars" of couplings. For instance, expanding the Einstein-Hilbert action to nth order in the fluctuation field results in n ``avatars" of the Newton coupling and cosmological constant, $\lambda_n$ and $G_n$. Tab.~\ref{tab:fluctab} lists fixed-point results for these ``avatars". We use the notation $\mu=-2\lambda_2$ and also provide the fluctuation field anomalous dimension $\eta_h$ and ghost anomalous dimension $\eta_c$. Where their full momentum dependence has been evaluated, as in \cite{Christiansen:2012rx,Christiansen:2015rva,Denz:2016qks}, the numbers provided refer to anomalous dimensions at vanishing momentum.
 ``Hybrid" calculations, which evaluate the anomalous dimensions of the fluctuation fields, but equate the background and fluctuation Newton couplings, $G_B$ and cosmological constants, $\Lambda_B$ are included. 
\begin{table}[!t]
\scriptsize
\begin{tabular}{c|c|c|c|c|c|c|c|c|c|c|c|c|c|c}
Ref. & gauge& regulator&bckr.&$\mu_{\ast}$ &$\lambda_3$& $G_{3\, \ast}$ & $G_{4\, \ast}$  & $\eta_h$ & $\eta_c$& ${\rm Re}\,\theta_1$ & ${\rm Re}\,\theta_2$ & ${\rm Re}\,\theta_3$\\\hline\hline
\cite{Groh:2010ta} &$\beta= \alpha=1$&Litim&sphere&$\Lambda_B=0.14$&- & $G_{B}=0.86$ &-  &-&-1.77&1.94 & 1.94 & - \\ \hline
\cite{Eichhorn:2010tb} &  $\beta=\alpha=0$& exp. &flat/sphere&$\Lambda_B=0.32$ &-&$G_{B}=0.29$ &- &- & -0.78& 2.03 &2.03 &- \\ \hline
\cite{Eichhorn:2010tb} &  $\beta=\alpha=1$& exp. &flat/sphere&$\Lambda_B=0.48$ &-&$G_{B}=0.18$ &- &- & -1.31& 1.39 &1.39 &- \\ \hline
\cite{Christiansen:2012rx} & $\beta=1,\alpha=0$& Litim&flat& -0.49 &-& 0.83 & -  &  0.5& -1.37&1.87&1.87&1.37\\\hline
\cite{Codello:2013fpa}& $\alpha=\beta=1$& Litim&flat& $\Lambda_B=-0.06$& -& $G_B=1.62$ &-&0.69 & -1.36&4.12&4.12&-\\\hline
\cite{Christiansen:2015rva}&$\beta=1,\alpha=0$& Litim&flat&-0.59 & 0.11&0.66 &-&$\eta_h(p^2)$& $\eta_c(p^2)$&1.4&1.4 &-14\\\hline
\cite{Denz:2016qks}&$\beta=1,\alpha=0$& Litim& flat&-0.45&0.12&0.83 &0.57&$\eta_h(p^2)$& $\eta_c(p^2)$&4.7&2.0&2.0\\\hline
\cite{Knorr:2017fus}&$\beta=1,\alpha=0$& Litim&curved&0.20&-0.008 &0.20&-&-&- &1.65 &1.65&-5.43\\\hline
\cite{Christiansen:2017bsy}&$\beta=1,\alpha=0$& Litim&curved&-0.38&-0.12&0.60&-&-&-&2.1&2.1&-3.5\\\hline
\end{tabular}
\caption{\label{tab:fluctab} Fixed-point results for fluctuation couplings. We caution that where several  ``avatars" of a coupling are present these are related by STIs. Accordingly not all critical exponents are physical.
}
\end{table}

The example of a background-deformed regularization for scalar field theory shows how the background dependence of the regulator can spoil the study of fixed-point results for the Wilson-Fisher fixed point \cite{Bridle:2013sra}.
 A symmetry identity, namely the shift Ward-identity, follows from background independence. It is structurally similar to the flow equation and relates the background-field-dependence on $\bar{\phi}$ and the fluctuation-field-dependence on $\varphi$ of the flowing action \cite{Reuter:1997gx,Litim:2002hj,Bridle:2013sra,Safari:2015dva},
\be
\frac{\delta\Gamma_k}{\delta \bar{\phi}}- \frac{\delta \Gamma_k}{\delta \varphi} = \frac{1}{2}{\rm Tr} \left[\left(\frac{\delta^2\Gamma_k}{\delta\varphi^2}+R_k[\bar{\phi}] \right)^{-1}\frac{\delta R_k[\bar{\phi}]}{\delta\bar{\phi}}\right].\label{eq:sWI}
\ee
Imposing the shift Ward-identity allows to recover background-independent results \cite{Bridle:2013sra}. In a similar spirit, studies imposing the shift Ward identity in background- approximations for gravity (where the analogue of Eq.~\eqref{eq:sWI} is supplemented by contributions from the gauge fixing sector) have been performed in \cite{Morris:2016spn,Percacci:2016arh,Labus:2016lkh,Ohta:2017dsq,Nieto:2017ddk}.

Dynamical triangulations are background-independent as there is no preferred configuration and even the foliation structure in CDTs appears to be dispensable \cite{Jordan:2013awa}. Therefore, establishing whether a universal continuum limit exists in the same universality class (i.e. with matching physical critical exponents) as FRG studies indicate,  tests background independence of asymptotically safe gravity.
 One can either approach this by the well-tested method of computer simulations, based on a Monte-Carlo approach, or explore tensor models (see Sec.~\ref{sec:RGQG}).

\subsubsection{The RG perspective on (discrete) quantum gravity}\label{sec:RGQG}
The use of RG ideas in quantum gravity has been gaining traction in various forms over the last few years. Interacting fixed points play a role in several different approaches, see, e.g., \cite{Dittrich:2014ala,Dittrich:2014mxa,Dittrich:2016tys,Bahr:2016hwc,Bahr:2017klw,Eichhorn:2013isa,Eichhorn:2017xhy,Benedetti:2014qsa,BenGeloun:2018ekd,Ambjorn:2014gsa} and references therein.  In particular, in models that introduce a discretization, RG tools enable searches for a universal continuum limit  encoded in RG fixed points. As one example, consider tensor models.  These are spacetime-free models which encode the gluing of fundamental building blocks of a triangulation in their combinatorics. They generate the sum over all simplicial pseudomanifolds (triangulations) through their Feynman-diagram expansion, thereby generalizing the success-story of matrix models \cite{DiFrancesco:1993cyw} to higher dimensions \cite{Ambjorn:1990ge,Godfrey:1990dt,Gross:1991hx}. A universal continuum limit might exist if the couplings are tuned to critical values while the tensor size $N$ is taken to infinity \cite{Gurau:2010ba}. This limit corresponds to a fixed point of an abstract, non-local RG flow set up in the tensor size $N$, \cite{Brezin:1992yc,Eichhorn:2013isa}. This coarse-graining flow goes from many degrees of freedom (large $N$), to fewer degrees of freedom (small $N$). It is background independent by making no reference to locality or spacetime. Therefore, if a viable fixed point, leading to a physically acceptable phase of spacetime (where the ``emergent" spacetime is four dimensional at large scales) can be identified, this provides an indication for  a universal continuum limit - i.e., asymptotic safety - in a background independent setting. In \cite{Eichhorn:2013isa} an  FRG approach was proposed for matrix models and generalized for tensor models in \cite{Eichhorn:2017xhy}, also triggering activity in related group field theories, e.g., \cite{Benedetti:2014qsa,BenGeloun:2018ekd}.

\subsubsection{Towards asymptotically safe phenomenology in astrophysics and cosmology}
As a candidate for a model of quantum spacetime, asymptotic safety should explain the structure of spacetime in the very early universe, see \cite{Bonanno:2017pkg} for a review and in those regions of black-hole spacetimes that contain classical curvature singularities.
Within the FRG language, the UV physics is encoded in the limit of the full effective action $\Gamma_{k \rightarrow 0}$ in which \emph{physical} scales, e.g.,  curvature scales, are taken to (trans)planckian values. External physical scales can act as an IR cutoff for quantum fluctuations, as is most easily seen for the external momenta in scattering processes. This motivates the use of "RG improvement" techniques that provide quantum-gravity ``inspired" models.  The RG-improvement is performed by upgrading all couplings to running couplings and subsequently identifying $k$, either at the level of the action, the equations of motion or the classical solutions. 
In settings with a high degree of symmetry and correspondingly a single physical scale, the identification is unique and dictated by dimensional arguments (e.g., $k^2 \sim R$ is the unique choice for a deSitter-type setting).``RG improved" results indicate dimensional reduction of the spectral dimension \cite{Lauscher:2005qz,Reuter:2011ah,Calcagni:2013vsa}, singularity resolution in black holes \cite{Bonanno:1998ye,Bonanno:2000ep,Bonanno:2006eu,Falls:2010he,Becker:2012js,Falls:2012nd,Koch:2013owa,Koch:2014cqa,Kofinas:2015sna,Pawlowski:2018swz,Adeifeoba:2018ydh}, finite entanglement entropy \cite{Pagani:2018mke} as well as an inflationary regime generated through quantum gravity effects \cite{Bonanno:2008xp,Bonanno:2010bt,Bonanno:2010mk,Reuter:2012xf,Kofinas:2016lcz}.

\section{Asymptotically safe quantum gravity and matter}
Our universe contains gravitational and matter degrees of freedom which are coupled to each other. Thus, to understand the quantum structure of spacetime in our universe it is neither necessary nor sufficient to show consistency of quantum- gravity models disregarding matter. This does not imply that quantum gravity must  be a unified theory of all interactions, or that it needs to contain matter as fundamental degrees of freedom. It simply means that at observationally accessible scales accessible, all  degrees of freedom, gravitational and matter, must be accounted for and their predicted dynamics compatible with observations.
As the Standard Model contains $N_S=4$ scalars, $N_V=12$ and $N_D=24$ fermions (including right-handed neutrinos) but there is only one metric, the microscopic gravitational dynamics might even be well-approximated by the dynamics obtained from an appropriate large $N_i$ approximation.

The measured Higgs mass of  $M_h\approx 125\, \rm GeV$ \cite{Aad:2012tfa,Chatrchyan:2012xdj}  lies within a narrow band where no new physics is required for the consistency of the Standard Model below the Planck scale. 
A Higgs mass higher than about 180 GeV \cite{Hambye:1996wb} leads to Landau-pole type behavior in the quartic coupling below the Planck scale. In the absence of higher-order terms in the Higgs potential, see, e.g.,\cite{Branchina:2013jra,Gies:2013fua,Eichhorn:2015kea})  the lower bound on the Higgs mass from absolute vacuum stability lies at $M_h=129\, \rm GeV$ in a three-loop study for $\Lambda_{\rm NP}=M_{\rm Pl}$, \cite{Bezrukov:2012sa}, rendering the electroweak vacuum metastable. As its lifetime exceeds the age of the universe \cite{EliasMiro:2011aa}, see \cite{Markkanen:2018pdo} for a review, the next scale of new physics for the Standard Model could be the Planck scale.
Such a ``desert"  provides an exciting opportunity for quantum gravity: The initial conditions for the RG flow of matter interactions are set by quantum gravity at the Planck scale. In the absence of the ``desert", new physics at intermediate scales could shield the quantum gravity scale from view. Conversely, in a ``desert"-like setting, there is a direct link between Planck-scale physics and electroweak-scale physics.

\subsection{Impact of quantum gravity on matter}
There are two effects of asymptotically safe gravity on matter in truncated FRG studies. Firstly, it generates nonzero fixed-point values for particular higher-order matter couplings, see Sec.~\ref{subsubsec:WGB}. Secondly, it impacts the scale dependence of the canonically marginal Standard Model couplings, see Sec.~\ref{subsubsec:linkmatters}. Both effects result in observational consistency constraints on the microscopic gravitational parameter space.

\subsubsection{Matter interacts in the presence of asymptotically safe quantum gravity}\label{subsubsec:WGB}
The interacting nature of the asymptotically safe gravitational dynamics percolates into the matter sector. There cannot be UV fixed point with \emph{all} matter interactions set to zero, \cite{Eichhorn:2011pc}\footnote{That four-fermion interactions are generated by quantum gravity fluctuations but remain finite implies that chiral symmetry, protecting the light fermions of the Standard Model, remains intact, see also \cite{Meibohm:2016mkp,Eichhorn:2017eht}. The effective background curvature in the UV can nevertheless break chiral symmetry \cite{Gies:2018jnv}.} \footnote{In $d \neq 4$, where specific matter models feature interacting fixed points, it is an intriguing question whether a new, combined universality class for matter and asymptotically safe gravity exists, see, e.g., \cite{Elizalde:1995ad,Percacci:2015wwa,Labus:2015ska}.}.
Interactions respecting the global symmetries of the kinetic terms for matter fields cannot be set to zero consistently \cite{Eichhorn:2017eht}. Finite contributions to their beta functions  are generated by gravitational fluctuations. These  prevent a free fixed point, as they are independent of the matter coupling and instead scale with the Newton coupling $G$ \cite{Eichhorn:2011pc,Eichhorn:2012va,Eichhorn:2016esv,Christiansen:2017gtg,Eichhorn:2017eht,Eichhorn:2017sok}. 
Thus, the free fixed point that exists in the limit of vanishing Newton coupling, $G \rightarrow 0$, is shifted to a finite  value, the shifted Gaussian fixed point (sGFP).
  \begin{figure}[!t]
  \centering
\includegraphics[width=0.5\linewidth]{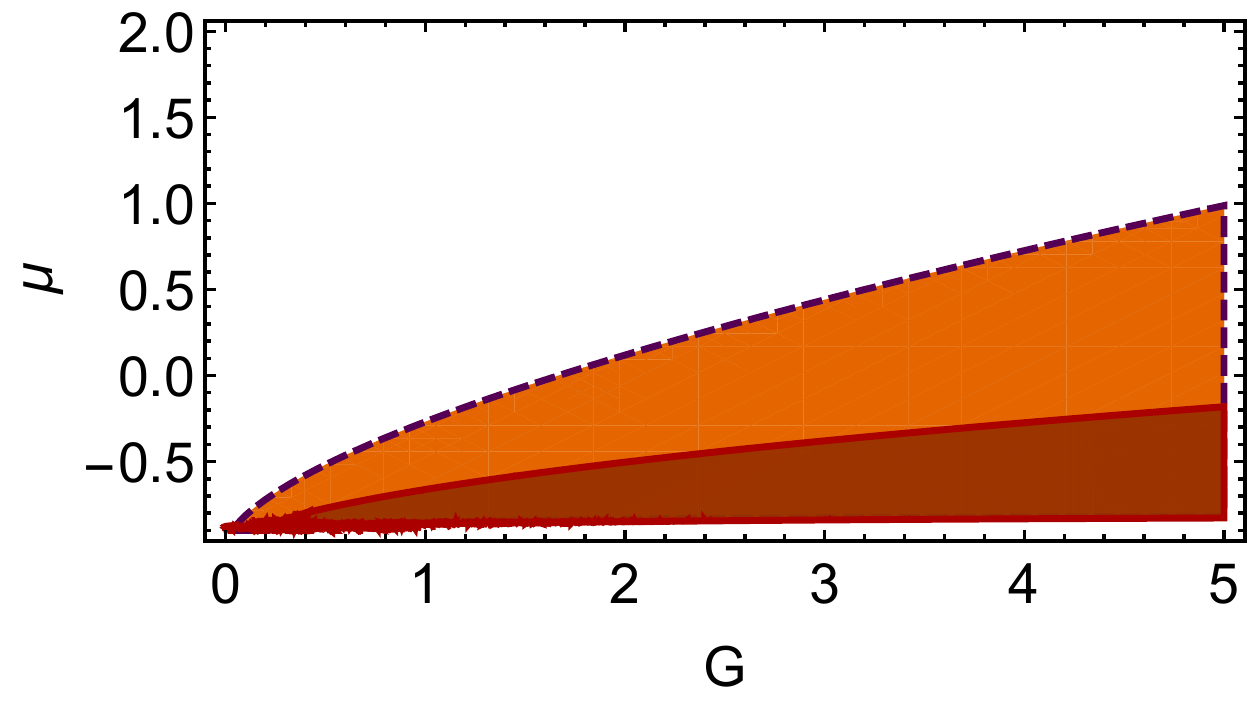}
\caption{\label{fig:wgb}The weak gravity bound in the $(G, \mu=-2\Lambda)$ plane for the Yang-Mills system (from \cite{Christiansen:2017gtg}) in orange, bounded by the dark dashed line, and the weak gravity bound in scalar-fermion systems (from \cite{Eichhorn:2016esv,Eichhorn:2017eht}) in dark red, bounded by the red continuous line, lie close to each other. 
}
\end{figure}
 Matter couplings $\bar{\chi}$ invariant under the global symmetries of the kinetic terms \footnote{ Notwithstanding arguments that suggest that quantum gravity should break global symmetries \cite{Kallosh:1995hi}, studies of the  FRG flow in truncations indicate the opposite result. This might be tied to the potential existence of black-hole remnants in asymptotic safety \cite{Bonanno:2000ep,Falls:2010he}. This implies that, e.g., Standard-Model couplings do not feature a contribution $\sim \#_2$, only a term $\sim \#_1$.} feature canonical dimensions  $d_{\rm \bar{\chi}}<0$ in $d=4$. Schematically, the FRG beta function reads
 \be
 \beta_{\chi} = -d_{\rm \bar{\chi}} \chi + \#_1 G_{\rm eff} \chi + \#_2\, G_{\rm eff}^2+ \#_3\, \chi^{\#}.
 \ee
 We focus on $\#=2$, see \cite{Eichhorn:2011pc,Eichhorn:2012va,Eichhorn:2016esv,Christiansen:2017gtg,Eichhorn:2017eht}.
$G_{\rm eff} = \frac{G}{1+\mu} - \frac{G}{(1+\mu^2)}$ parameterizes the effective strength of gravity fluctuations in the Einstein-Hilbert truncation, see \cite{Eichhorn:2017eht} for higher-order terms. The fixed points are 
 \be
 \chi_{1/2\, \ast}= \frac{d-\#_1\, G_{\rm eff} \pm \sqrt{-4 \#_2\, \#_3\, G_{\rm eff}^2 + (\#_1 G_{\rm eff}-d_{\rm \bar{\chi}})^2}}{2\,\#_3},
 \ee 
such that $\chi_{1\, \ast}$ is the sGFP. For ${\rm sign}\#_3= {\rm sign}\#_2$, these two fixed points collide at 
 \be
 G_{\rm \, eff,\, crit}= \frac{d_{\rm \bar{\chi}}}{\#_1 - 2 \sqrt{\#_2\, \#_3}}.
 \ee
Beyond, the sGFP is complex, thus $G> G_{\rm eff, \, crit}$ is inconsistent. As $G_{\rm eff}$ measures the effective strength of gravity fluctuations, $G_{\rm eff, \, crit}$ marks the (truncation dependent) weak-gravity bound.
Once gravitational fluctuations exceed this  bound, cf.~Fig.~\ref{fig:wgb}, they trigger novel divergences in matter couplings, restricting the viable microscopic parameter space to the remaining region.
As the induced matter couplings are canonically irrelevant, they are power-law suppressed below the Planck scale and presumably irrelevant for particle physics at lower scales.

\subsubsection{A link that could matter: From the Planck scale to the electroweak scale}\label{subsubsec:linkmatters}
Asymptotically safe gravity could uniquely fix the values of marginally irrelevant Standard Model (SM)-couplings (Abelian gauge couplings, Yukawas, Higgs quartic) at the Planck scale. This might allow to confront asymptotic safety with observations, since those couplings run logarithmically below the Planck scale, retaining a ``memory" of their Planck-scale values.  

For the marginal SM couplings $g_{\rm SM}$, the quantum-gravity contribution to  $\beta_{g_{\rm SM}}$ is linear in $g_{\rm SM}$, as  the gravitational RG flow cannot generate the SM interactions once they are set to zero due to their distinct symmetry structure \cite{Eichhorn:2017eht}. Technically, this is encoded in the diagrams underlying the  FRG flow, see, e.g., \cite{Christiansen:2017gtg,Eichhorn:2017lry,Eichhorn:2016esv,Eichhorn:2017eht,Eichhorn:2017als}.
Hence, the quantum-gravity contribution is
\be
\beta_{g_{\rm SM}}\Big|_{\rm grav} = - f_{g_{\rm SM}}\,g_{\rm SM}, 
\ee
where $f_{g_{\rm SM}} \sim G$ is the contribution of metric fluctuations to the corresponding interaction vertex and additionally contains the gravity contribution to the anomalous dimensions of the matter fields.
This contribution acts like a scaling dimension, i.e., like an effective change in spacetime dimensionality. 
For canonically irrelevant couplings, a UV completion requires $f_{g_{\rm SM}}>0$, resembling an effective \emph{dimensional reduction}. It is unclear whether and how this  fits with other indications for dimensional reduction in quantum gravity \cite{Carlip:2017eud}.

Asymptotic freedom in non-Abelian gauge theories is a key cornerstone in the construction of the SM. This property could persist, as 
\be
\beta_g = -f_g \,g- \#_g\, g^3,...\label{eq:betagnA}
\ee
for gauge couplings $g$, where $f_g\geq0$ holds in all FRG studies to date \cite{Daum:2009dn,Folkerts:2011jz,Harst:2011zx,Christiansen:2017gtg,Eichhorn:2017lry,Christiansen:2017cxa}.
$\#_g$ depends on the gauge group and matter content while the gravity contribution is blind to the internal index structure and accordingly gauge-group independent.  Additional gravity contributions are indirect ones, arising through quantum-gravity-induced higher-order interactions which couple into the flow of the gauge coupling, \cite{Christiansen:2017gtg} (note that the sign of the $w$-term in $\eta$ is incorrect; accordingly this indirect contribution strengthens asymptotic freedom.)\\
The non-universality of beta functions, setting in at three loops for dimensionless couplings, starts at leading order for dimensionful couplings. Hence, the gravity contributions to beta functions in approximations differ in different schemes, see    \cite{Robinson:2005fj,Pietrykowski:2006xy,Toms:2007sk,Ebert:2007gf,Toms:2008dq,Ebert:2008ux,Toms:2009vd,Toms:2010vy,Felipe:2011rs,Narain:2013eea,Ellis:2010rw,Gonzalez-Martin:2017bvw,Toms:2011zza,Rodigast:2009zj,Mackay:2009cf,Pietrykowski:2012nc,Anber:2010uj,Anber:2011ut} for perturbative studies. At the level of observables, such dependences must cancel. The same physics is encoded in different ways in distinct schemes. As the FRG is applicable to settings with dimensionful couplings (including a multitude of extensively probed universality classes in statistical physics), one could argue that it is well-suited to explore quantum gravity in simpler approximations.
The non-universality of the gravity-contribution is reflected in the regulator-dependence of $f_g$ in truncations: Within a background-field study, $f_g =\frac{6}{\pi}\Phi_1^1(0)$ \cite{Daum:2009dn}, where $\Phi_1^1(0)>0$ always holds, but the value  depends on the choice of regulator, e.g., $\Phi_1^1(0)=1$ for the Litim-cutoff and $\Phi_1^1(0)=\pi^2/6$ for the exponential cutoff. This dependence is expected to cancel against regulator-dependence of gravitational fixed-point values (at least at the level of physical observables).
   
 \begin{figure}[!]
\centering
\includegraphics[width=0.5\linewidth]{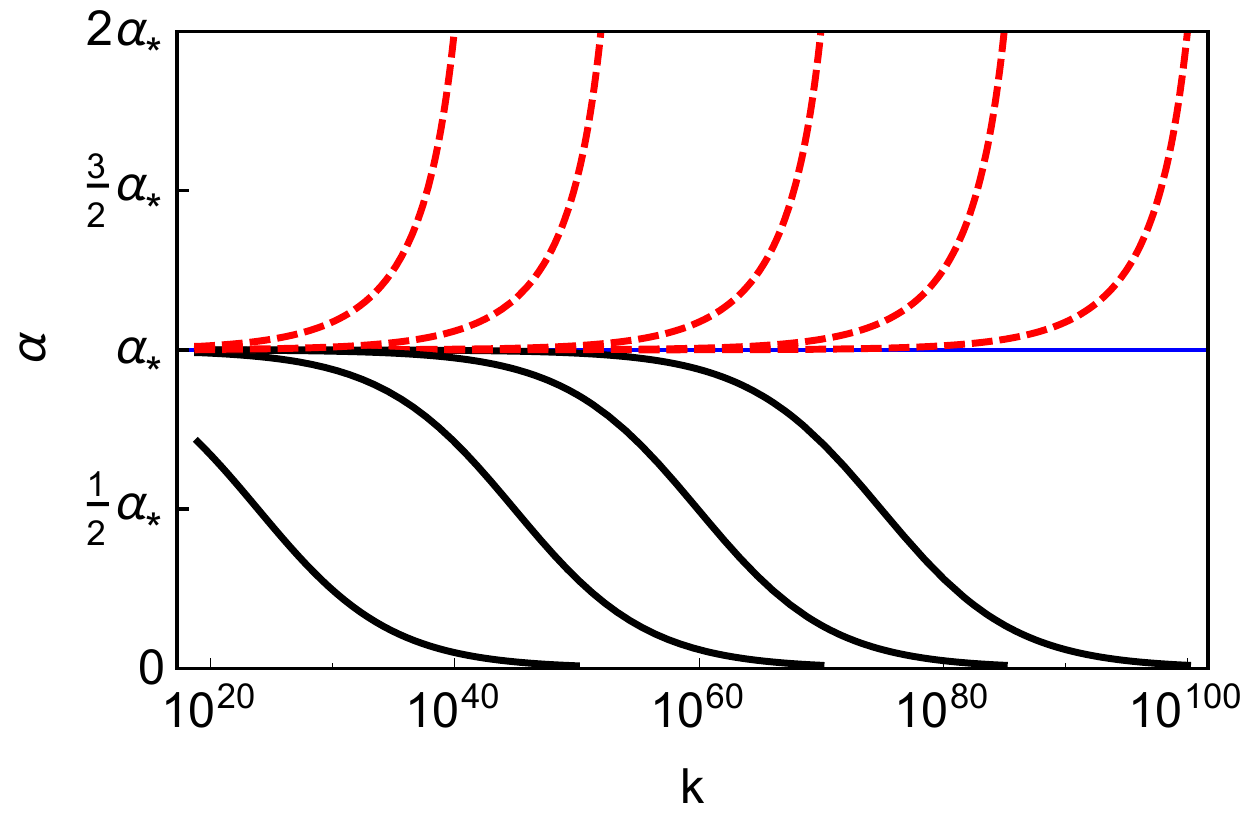}
\caption{\label{fig:alphabound} The transplanckian RG flow for $\alpha=g_Y^2/(4 \pi)$ described by Eq.~\eqref{eq:betagy} features trajectories emanating from the free fixed point (black, continuous line), which approach the interacting fixed point at $\alpha_{\ast}$. One unique trajectories (blue, thin line) is the fixed-point trajectory for the interacting fixed point. UV unsafe trajectories are pulled towards the IR fixed point as well (red, dashed lines).
Adapted from \cite{Eichhorn:2017muy}.}
\end{figure}

For the Abelian gauge coupling the free fixed point is IR attractive in the absence of gravity, such that the observation of a nonvanishing Abelian gauge coupling in the IR presumably prevents an asymptotically free UV completion of the SM \cite{GellMann:1954fq}. 
The quantum gravity contribution is the same as in the non-Abelian case (cf.~Eq.~\eqref{eq:betagnA}), thus
\be
\beta_{g_Y}=- f_g\, g_Y + \frac{41}{6}\frac{g_Y^3}{16\pi^2}+...\label{eq:betagy}
\ee
Fixed points of Eq.~\eqref{eq:betagy} lie at 
\be
g_{Y, \ast\, 1}=0, \quad g_{Y, \ast\, 2}=\sqrt{\frac{f_g\, 6\cdot 16\pi^2}{41}}.
\ee
The first is IR repulsive, the second IR attractive. If it lies at small enough values, then higher-order terms remain negligible and Eq.~\eqref{eq:betagy} suffices to  analyze the consequences. According to Eq.~\eqref{eq:betagy}, the IR repulsive fixed point at $g_{Y\, \ast\,1}=0$ can be connected to a range of values for $g_Y$ at the Planck scale \footnote{Couplings that are asymptotically free, not asymptotically safe, already run at transplanckian scales.}. However, no values above an upper bound, $g_{Y}=g_{Y\, \ast\, 2}$, can be reached, as $g_{Y\, \ast\,2}$ is IR attractive, cf.~Fig.~\ref{fig:alphabound}. 
Only one unique trajectory emanates from $g_{Y\, \ast\, 2}$. Along this trajectory, $g_Y(k)=\rm const$ until quantum-gravity contributions switch off below the Planck scale, where $f_g$ quickly drops to tiny values and SM fields drive the flow. Unlike in the SM without the gravity-extension, the initial condition for the RG flow of $g_Y$ is fixed at the Planck scale. Testing whether this results in an observationally viable value  at the electroweak scale constitutes a strong observational test of the model. It also highlights that confronting quantum gravity with observations might be possible without reaching Planckian energies.

\begin{figure}[!t]
\includegraphics[width=0.45\linewidth]{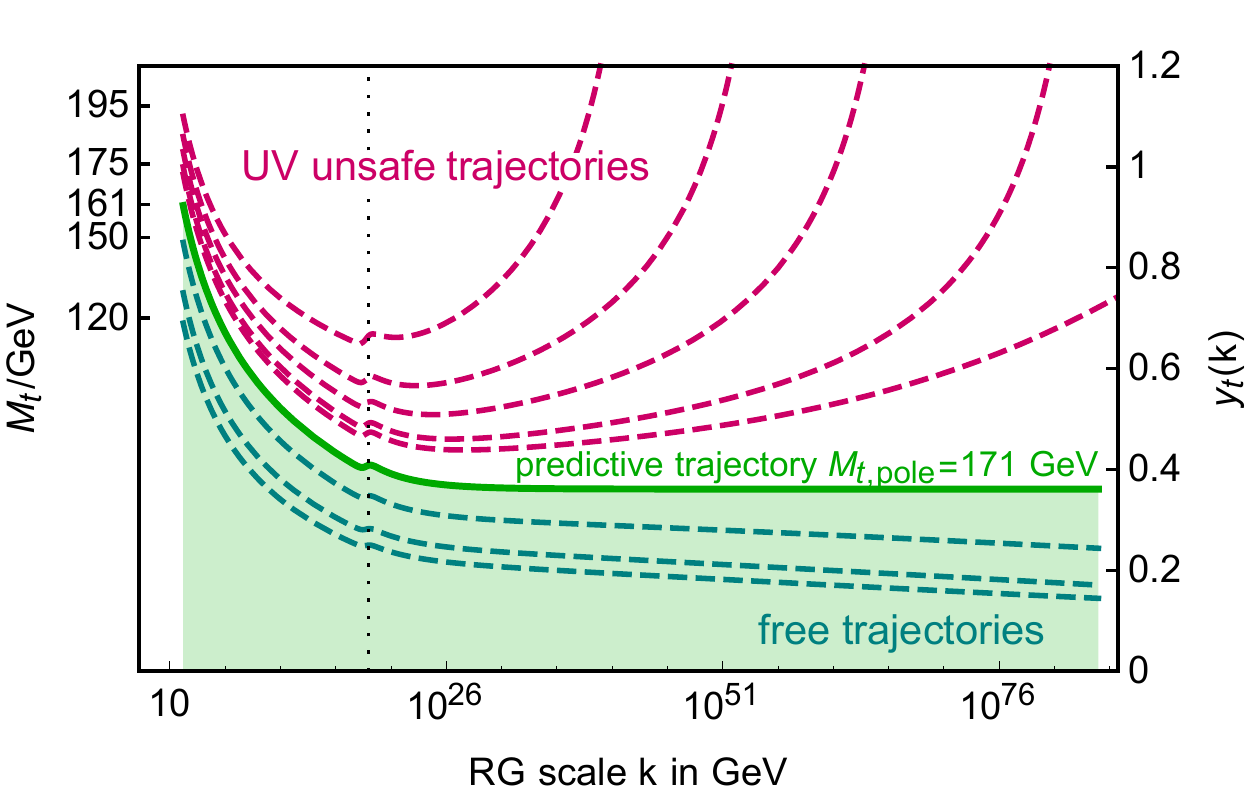}\quad \includegraphics[width=0.45\linewidth]{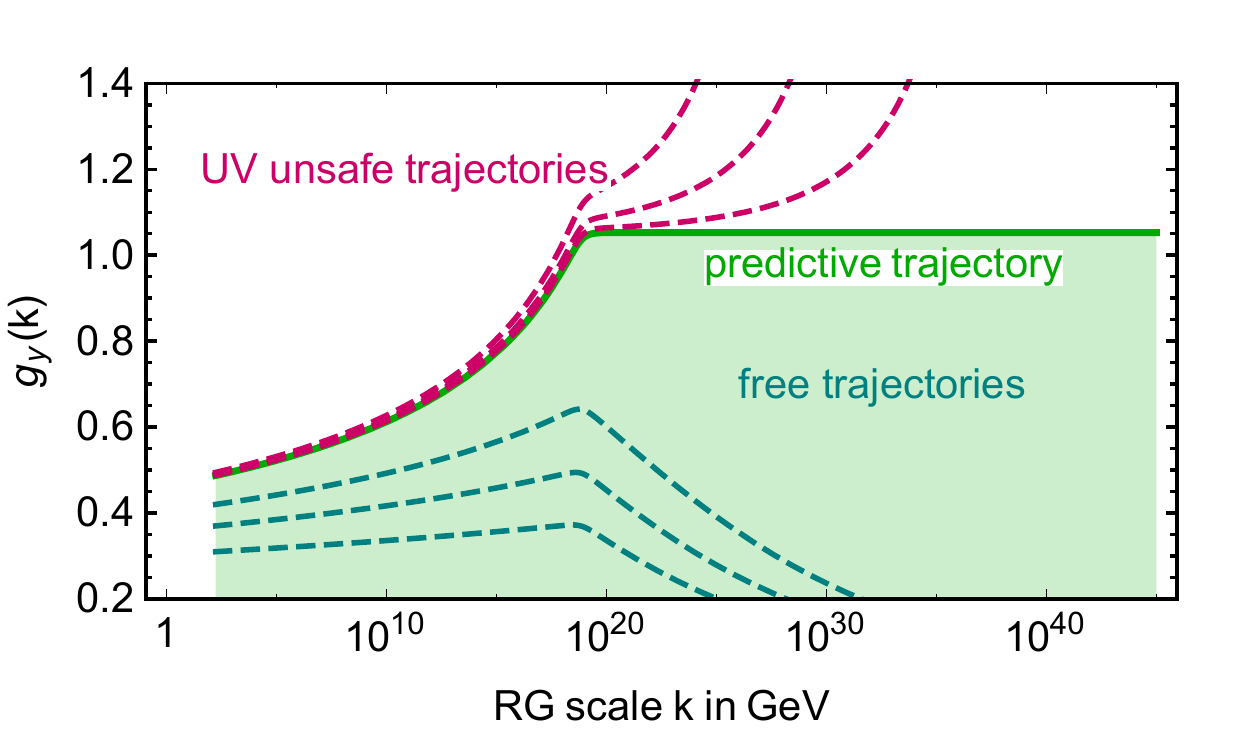}
\caption{\label{fig:upperbounds} From \cite{Eichhorn:2017ylw} and \cite{Eichhorn:2017lry}.}
\end{figure}

The  fixed-point structure underlying such ``retrodictions"  was found for the Abelian gauge coupling in \cite{Harst:2011zx}, further explored in \cite{Eichhorn:2017lry},  cf.~right panel in Fig.~\ref{fig:upperbounds} and extended to a GUT setting in \cite{Eichhorn:2017muy}. 

In \cite{Zanusso:2009bs,Vacca:2010mj,Oda:2015sma,Eichhorn:2016esv,Eichhorn:2017eht,Hamada:2017rvn}, the gravity-contribution $f_y$ to the Yukawa sector was calculated. Using beta functions of the form
\bea
\label{eq:betays}
\beta_{y_{t \,(b)}} =
\frac{y_{t \,(b)}}{16\,\pi^2}\;\left(\frac{3 y_{b \,(t)}^2}{2}+ \frac{9 y_{t \,(b)}^2}{2}  - \frac{9}{4}g_2^2 -8 g_3^2 \right) - f_y\, y_{t \,(b)}  -\frac{3 y_{t \,(b)}}{16\,\pi^2}\left(\frac{1}{36}+Y_{t \,(b)}^2\right)g_Y^2,
\eea
 for the quarks of the third generation, with $Y_t=2/3$, $Y_b=-1/3$, supplemented by the assumption that the gauge sector of the SM is asymptotically free, and gravitational fixed-point values from a background-approximation results in a uniquely fixed top mass of about 170 GeV  \cite{Eichhorn:2017ylw}, cf.~left panel of Fig.~\ref{fig:upperbounds}.\\
Intriguingly, the SM  beta functions with gravity in the approximation defined by Eq.~\eqref{eq:betagy} and Eq.~\eqref{eq:betays} also admit an interacting fixed point such that the top Yukawa, bottom Yukawa and Abelian gauge coupling are fixed uniquely. They reach IR values in the vicinity of the observed ones, if the two gravity contributions $f_g$ and $f_y$ take appropriate values \cite{Eichhorn:2018whv}. In this scenario, the difference between top mass and bottom mass is generated through an interacting fixed point induced by gravity due to their different charges.

\begin{figure}[!t]
\includegraphics[width=0.45\linewidth]{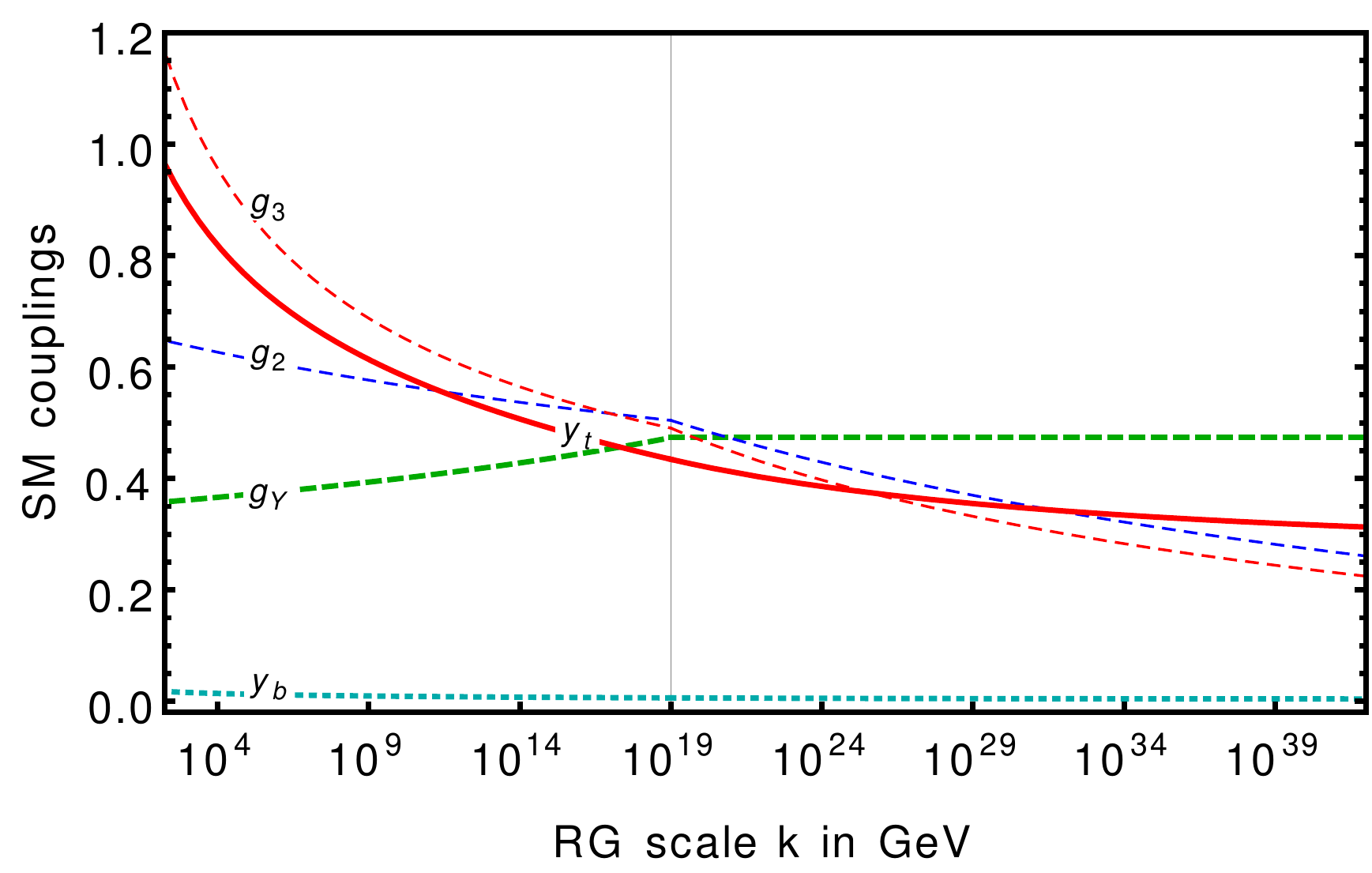}\quad \includegraphics[width=0.45\linewidth]{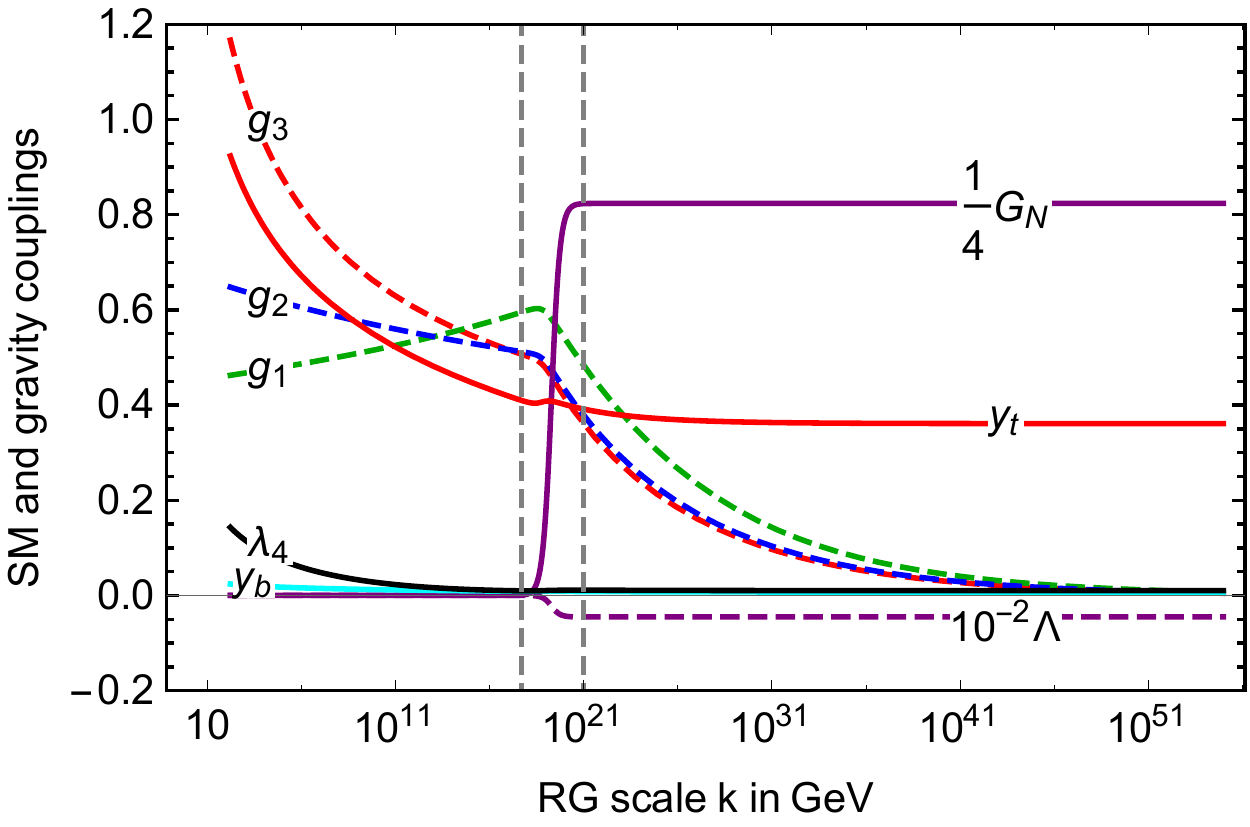}
\caption{\label{fig:runningSMextensions} Both panels: RG flows in an approximation as in Eq.~\eqref{eq:betagy}, see \cite{Eichhorn:2018whv} and \cite{Eichhorn:2017ylw} for details. Left panel: Flow of gauge couplings and top and bottom Yukawa with quantum-gravity parameterized by $f_g=9.8\cdot 10^{-3}$ and $f_y=1.13\cdot 10^{-4}$ above the Planck scale and $f_g=0=f_y$ below the Planck scale as in \cite{Eichhorn:2018whv}.
Right panel: Standard-Model RG flow including running gravitational couplings as in \cite{Dona:2013qba} and is taken from \cite{Eichhorn:2017ylw}.}
\end{figure}

The fixed-point structure could be simpler in the scalar sector.
Asymptotically safe quantum gravity flattens the Higgs potential: If all other SM couplings are asymptotically free, a fixed point at vanishing Higgs potential 
exists in line with intact shift-symmetry \cite{Eichhorn:2017eht}.  It is IR attractive \cite{Narain:2009fy, Percacci:2015wwa,Labus:2015ska,Oda:2015sma,Hamada:2017rvn,Eichhorn:2017als}. This extends to the Higgs portal coupling to scalar dark matter \cite{Eichhorn:2017als}. Taking the corresponding fixed-point values $\lambda_{h\ast}=0$ (for the Higgs quartic) and $\lambda_{h\chi\, \ast}=0$ (for the Higgs portal coupling) as initial conditions for the RG flow at the Planck scale, and setting all SM couplings to their observationally preferred Planck-scale values, one reaches a Higgs mass in the vicinity of the observed value, while the Higgs portal coupling remains zero at all scales. The first is a prediction \cite{Shaposhnikov:2009pv} put forward before the discovery of the Higgs at the LHC \cite{Aad:2012tfa,Chatrchyan:2012xdj}, see also \cite{Bezrukov:2012sa}. The second appears to be consistent with the non-detection of a scalar Higgs portal through direct searches \cite{Athron:2017kgt,Aprile:2018dbl}. 

   ``Retrodictions" of SM couplings could be a much more generic consequence of quantum gravity than just of asymptotic safety as discussed in~Sec.~\ref{sec:ASnonfund}.

\subsection{Impact of matter on quantum gravity -- backreaction matters?}
The impact of quantum fluctuations of matter on the gravitational fixed point has been studied in simple truncations. The corresponding theory space also contains non-minimal matter-curvature couplings, \cite{Narain:2009fy,Percacci:2015wwa,Eichhorn:2017sok,Eichhorn:2016vvy,Eichhorn:2017sok}.\\
Matter fields deform the gravitational fixed point in truncations. Adding a small number of matter fields leads to the continued existence of a viable interacting fixed point. At larger number of matter fields, there are indications that further extensions of the truncation could be required \cite{Meibohm:2015twa,Eichhorn:2018akn}.

Assuming that asymptotic safety in gravity is driven by antiscreening metric fluctuations inducing a fixed point in the Newton coupling, the matter contribution to $\beta_G$ is critical.  Specifically, 
\be
\beta_G\Big|_{\rm matter} = N_S\, G^2\, a_S + N_D\, G^2\, a_D+N_V\, G^2\, a_V,
\ee
where $a_S>0$ \cite{Dona:2013qba, Meibohm:2015twa,Labus:2015ska,Percacci:2015wwa,Dona:2015tnf,Biemans:2017zca,Alkofer:2018fxj,Eichhorn:2018akn}, agreeing with perturbative studies for $d=2+\epsilon$ dimensions \cite{Christensen:1978sc} and studies of the one-loop effective action using heat-kernel techniques \cite{Kabat:1995eq,Larsen:1995ax}. Similarly, fermions screen the Newton coupling \footnote{For the background Newton coupling, this is more subtle: Choosing to impose the regulator on the spectrum of $\nabla^2$, or on $\slashed{\nabla}^2 = \nabla^2-R/4$ results in a different sign of the fermionic contribution to the running of $G$ \cite{Dona:2012am}, see also \cite{Alkofer:2018fxj,Alkofer:2018baq}. This highlights that the (unphysical) background-field dependence of the regulator can alter results in the background approximation in simple truncations, suggesting the need for a fluctuation calculation. The first choice agrees with the result from fluctuations calculations.}, $a_D>0$ \cite{Dona:2013qba,Meibohm:2015twa,Eichhorn:2018ydy}, in agreement with perturbative studies \cite{Kabat:1995eq,Larsen:1995ax}. For vectors, $a_V<0$ \cite{Dona:2013qba,Christiansen:2017cxa,Biemans:2017zca,Alkofer:2018fxj,Eichhorn:2018ydy}, also found with perturbative techniques \cite{Kabat:1995eq,Larsen:1995ax}. Background and fluctuation results are in agreement on this result (for fluctuation results, it is crucial to include the anomalous dimensions \cite{Dona:2015tnf,Eichhorn:2018akn}). 

A strong indication for (near-perturbative) asymptotic safety in matter-gravity systems comes from a comparison \cite{Eichhorn:2018akn,Eichhorn:2018ydy}  of distinct ``avatars" of the Newton coupling  \cite{Dona:2015tnf}. It can be read off from the three-graviton vertex as well as gravity-matter vertices, just like the gauge coupling in gauge theories. For a dimensionless gauge coupling in the perturbative regime, two-loop universality equates the different avatars. Beyond perturbation theory, the Slavnov-Taylor-identities relating the avatars become nontrivial. Simply put, the stronger quantum effects are, the less trivial are the relation of classically equal couplings. \cite{Eichhorn:2018akn,Eichhorn:2018ydy} observe an effective universality of distinct avatars of the Newton coupling, which agree within an estimate of the systematic truncation error. This signals a near-perturbative nature of asymptotically safe gravity. Further, the delicate cancellations required between different contributions to the beta functions in order to achieve effective universality strongly point towards a physical fixed point instead of a truncation artifact.

\section{Outlook}
Asymptotically safe models are of inherent theoretical interest when it comes to a comprehensive understanding of fundamental quantum field theories. Exciting progress in the last few years even hints at a possibility of asymptotically safe extensions of the Standard Model -- with or without gravity. In quantum gravity, the idea of asymptotic safety resonates with a wider effort to analyze quantum spacetime from a Renormalization Group point of view.
Hence, the many intriguing open questions that remain to be answered in this area appear worth tackling, and the new (asymptotically safe) perspective on high-energy physics is exciting and potentially useful  to explore. \newline\\

\emph{Acknowledgements: 
I would like to thank all participants and speakers of the workshop ``Asymptotic safety in a dark universe" at Perimeter Institute in June 2018, the conference ``Quantum spacetime and the Renormalization Group" in Bad Honnef in June 2018 and the workshop on ``Quantum fields -- from fundamental question to phenomenological applications" at MITP in September 2018 for stimulating discussions, and in particular acknowledge helpful discussions with L.~Freidel, D.~Litim, J.~Pawlowski, M.~Reuter, F.~Sannino and C.~Wetterich that have informed this review. 
It is a particular pleasure to thank all current and past members of my research group for successful collaborations, inspiring discussions and for joining in the (only asymptotically safe) quest for quantum gravity.
I acknowledge support by the DFG under grant no.~Ei/1037-1.}

\end{document}